\documentclass[a4paper,fleqn,usenatbib]{mnras}
\usepackage{newtxtext,newtxmath}

\usepackage[T1]{fontenc}
\usepackage{ae,aecompl}

\usepackage{graphicx}	
\usepackage{amsmath}	
\usepackage{amssymb}	

\title[Gas-rich galaxies at $z \sim 5$]{Early metal enrichment of gas-rich galaxies at $z \sim 5$}

\author[S. Poudel et al.]{
Suraj Poudel,$^{1}$\thanks{E-mail: spoudel@email.sc.edu}
Varsha P. Kulkarni,$^{1}$
Sean Morrison,$^{2}$
C\'eline P\'eroux,$^{2}$
\newauthor
Debopam Som,$^{2}$
Hadi Rahmani$^{2}$
and Samuel Quiret$^{2}$
\\
$^{1}$University of South Carolina, Dept. of Physics and Astronomy, Columbia, Sc., 29208, USA\\
$^{2}$Laboratoire d'Astrophysique de Marseille, UMR 7326, F-13388, Marseille
}

\date{Accepted XXX. Received YYY; in original form ZZZ}

\pubyear{2017}

\begin{document}
\label{firstpage}
\pagerange{\pageref{firstpage}--\pageref{lastpage}}
\maketitle

\begin{abstract}
We present abundance measurements of elements O, C, Si and Fe for three gas-rich galaxies at z$\sim$5 using observations from the Very Large Telescope (VLT) and the Keck telescope in order to better constrain the early chemical enrichment of gas-rich galaxies. These galaxies show strong Lyman-$\alpha$ absorption in the spectra of background quasars, with neutral hydrogen column densities log $N_{\rm H\, I}$(cm$^{-2}$) = 20.10$\pm$0.15, 20.10$\pm$0.15, and 20.80$\pm$0.15. Using the undepleted element O, we find the metallicities [O/H] to be in the range of $-2.51$ to $-2.05$ dex. Our study has doubled the existing sample of measurements of undepleted elements at $z > 4.5$. Combining our measurements with those from the literature, we find that the $N_{\rm H\, I}$-weighted mean metallicity of z$\sim$5 absorbers is consistent with the prediction based on $z < 4.5$ DLAs. Thus, we find  no significant evidence of a sudden drop in metallicity at $z>4.7$ as reported by some prior studies. We also determine the extent of dust depletion using a combination of both the volatile element O and the refractory elements Si and/or Fe. Some of the absorbers show evidence of depletion of elements on dust grains, e.g. low [Si/O] or [Fe/O]. The relative abundances of these absorbers along with other z$\sim$5 absorbers from the literature show some peculiarities, e.g. low [C/O] in several absorbers and high [Si/O] in one absorber. Finally, we find that the metallicity vs. velocity  dispersion relation of z$\sim$5 absorbers may be different from that of lower-redshift absorbers.
\end{abstract}

\begin{keywords}
ISM: abundances -- galaxies: high-redshift -- quasars: absorption lines
\end{keywords}



\section{Introduction}
Understanding the early phases of metal and dust production in the universe is fundamental to many areas of astrophysics. The epoch $z\ga5$ spans roughly the first 1 billion years after the Big Bang. This exciting period  included the emergence of population III stars and the early generations of population II stars, which injected the surrounding medium with metals. Thus, one may expect to see rapid metallicity evolution and peculiar element abundance patterns in the interstallar medium (ISM) of $z\ga5$ galaxies, reflecting outputs of early stars \citep*[e.g.][]{Kulkarni et al. 2013, Maio 2015}.\\

Powerful probes of metals in distant galaxies are provided by the damped Lyman-alpha (DLA) and sub-damped Lyman-alpha (sub-DLA) absorbers in background quasar spectra, which have strong neutral hydrogen column densities (log $N_{\rm H\, I} \ge 20.3$ and 19.0 $\le$ log $N_{\rm H\, I} < 20.3$, respectively). This absorption-line technique samples galaxies simply by gas cross-section, independent of their brightness, and is not limited to only the bright or actively star-forming galaxies probed by flux-limited imaging surveys. DLAs and sub-DLAs  dominate the neutral gas mass density in the universe \citep*[e.g.][]{Zafar et al. 2013, Popping et al. 2014}. Observations of high-z DLAs can thus constrain galaxy evolution models \citep*[e.g.][]{Dave et al. 2011, Fumagalli et al. 2011, Cen 2012}. Furthermore, relative element abundances in DLAs can in principle be measured well into the reionization epoch, and can constrain the nucleosynthesis and initial mass function of population III stars \citep[e.g.,][]{Kulkarni et al. 2013}.\\

Observations of DLAs at $0 < z < 4.5$ indicate that their metallicity decreases at a modest rate of about 0.2 dex per unit redshift \citep[e.g.,][]{Prochaska et al. 2003a, Kulkarni et al. 2005, Kulkarni et al. 2007, Som et al. 2013, Som et al. 2015, Quiret et al. 2016, Rafelski et al. 2012, Jorgenson et al. 2013}. Remarkably, this rate of DLA metallicity evolution is similar to the rate of metallicity growth inferred from comparing the mass-metallicity relation (MZR) of low-mass star-forming galaxies at $z\sim0$ and $z\sim2$ \citep[e.g.,][]{Tremonti et al. 2004, Erb et al. 2006,  Henry et al. 2013}. To understand evolution of DLAs and their relation to star-forming galaxies at earlier cosmic epochs, it is necessary to extend the redshift baseline further to $z\ga5$. This is only now becoming feasible with the discovery of a number of DLAs with $z\ga5$ in the Sloan Digital Sky Survey (SDSS) \citep[e.g.,][]{Noterdaeme et al. 2012}.\\

\citet{Rafelski et al. 2012, Rafelski et al. 2014} reported a sudden drop in DLA metallicity at z>4.7 which, if real, may suggest changes in gas enrichment processes. However, this relatively small z>4.7 sample is based almost entirely on the elements Si and Fe which show significant depletion on dust grains in the interstellar medium (ISM) of the Milky Way \citep[e.g.][]{Jenkins 2009}. Indeed, Si and Fe are observed widely to be depleted in DLAs as well \citep[e.g.,][]{De Cia et al. 2016}. If these elements are depleted in the z>4.7 DLAs, the true metallicities could be higher, and the drop in metallicity at z>4.7 may be smaller. Dust depletion needs to be considered at $z\sim5$, given the evidence for dust in high-z galaxies, e.g., from $z\sim5$ sub-mm galaxies, and from dust evolution models \citep*[e.g.][]{Walter et al. 2012, Casey et al. 2014}. Volatile elements like O, S show much weaker depletions than Si, Fe in warm and cool Milky Way (MW) ISM and in DLAs at $z < 4.2$ \citep*[e.g.][]{Savage 1996, Jenkins 2009, De Cia et al. 2016}. Absorption lines of undepleted elements such as S or O thus offer more accurate metallicity indicators, but often lie in the Lyman$-\alpha$ forest. Therefore, current samples contain only a handful of O and S measurements at $z>4.5$ including only one for a sub-DLA  at $z=4.977$ \citep[]{Morrison et al. 2016}.\\

In this paper, we report element abundance measurements, including those for undepleted elements, in three absorbers at redshifts z > 4.8. We combine these results with those for other $z > 4.5$ absorbers, and compare them with lower-redshift measurements to obtain improved constraints on the early chemical enrichment history of galaxies. We also examine relative element abundances. Section~\ref{sec:obs} describes the observations and data reduction. Section~\ref{sec:result} discusses the results for individual absorbers. Section~\ref{sec:dust} presents a discussion of our results. Finally, section~\ref{sec:conclusion} summarizes our conclusions.
 

\section{Observations and data reduction}
\label{sec:obs}
The spectrum of the quasar Q0824+1302 ($m_{i}  = 19.95$) shows two absorbers at $z=4.829$ and $z=4.809$. The spectrum of the quasar Q0231-0728 ($m_{i} = 19.48$) shows an absorber at $z=5.335$. Table~\ref{tab:sum_obs} summarizes the properties of these targets and details of the observations. For Q0824+1302, three exposures of 2400 s each were obtained with VLT X-Shooter on 2014 December 15 as a part of the program 094.A-0793(A) (PI. S. Quiret). The X-Shooter Common Pipeline Library \citep{Goldoni 2006} release 6.5.1\footnote{http://www.eso.org/sci/facilities/paranal/instruments/xshooter/} was used to reduce the science raw images and to produce the final 2-D spectra. All of our science data were taken in the \textsc{nodding} mode, so a standard procedure of data reduction was used as follows. An initial guess was first computed for the wavelength solution and the position of the centre and the edges of the orders. Then the accurate position of the centre of each order was traced, and this step was followed by generating the master flat frame out of five individual lamp flat exposures. Next a 2-D wavelength solution was found, and it was modified by applying a flexure correction to correct for the shifts that can be of the order of the size of a pixel. Finally, having generated the required calibration tables, each pair of science frames was reduced to obtain the flat-fielded, wavelength-calibrated and sky-subtracted 2-D spectrum. The 1-D flux of the quasar was extracted using a spectral point spread function (SPSF) subtraction as described in \citet{Rahmani 2016}. In summary, the quasar's PSF was modeled using a Moffat function whose FHWM and centre were smooth functions of wavelength. The light profile was then integrated at each wavelength to obtain the quasar's flux. The data for Q0231-0728 were downloaded from the Keck Observatory Archive (KOA), and consisted of observations obtained with the Keck Echellette Spectrograph and Imager (ESI) as part of program C70E (PI S. Djorgovski). Individual spectra were extracted and normalised by fitting a cubic spline polynomial, typically of the sixth order, using the CONTINUUM task in IRAF\footnote{IRAF is distributed by the National Optical Astronomy Observatory, which is operated by the Association of Universities for Research in Astronomy, Inc., under cooperative agreement with the National Science Foundation.} and were combined thereafter. 
\subsection{Absorption Line Measurements and Column Density Determinations}
Column densities of H I and the heavy elements were determined using the Voigt profile fitting program VPFIT\footnote{https://www.ast.cam.ac.uk/rfc/vpfit.html} (version 10.0). All the available lines of O I, C II, Si II and Fe II were fitted with the effective Doppler b-parameters and redshifts of the corresponding velocity components tied together. The abundances of each element X were calculated using the relation, [X/H] = log ($N_{X}/N_{\rm H\, I}$) - log (X/H)$_\odot$, where $N_{X}$ is the column density of element X, $N_{\rm H\, I}$ is the neutral hydrogen column density and log $(X/H)_{\odot}$ is the solar abundance of the element X. The atomic data required for the measurement of different lines were adopted from \citet{Cashman et al. 2017}, and solar abundances were adopted from \citet{Asplund et al. 2009}. The metal lines O I $\lambda$1302, C II $\lambda$1334, Si II $\lambda$1304, Si II $\lambda$1260, Si II $\lambda$1527, Fe II $\lambda$$\lambda$ 2587, 2600 used for the determination of the abundances were outside the Lyman$-\alpha$ forest, which allows us to calculate the column densities reliably. 

We note that line saturation can be an issue in determination of the metal line column densities \citep[e.g.][]{Penprase et al. 2010}. At the medium resolution ($\sim 34-56$ km s$^{-1}$) of spectrographs such as X-shooter and ESI (used by us and other studies of $z \sim 5$ absorbers), it can be difficult to fully assess hidden saturation from much narrower components in single lines. Therefore, wherever possible, we have made use of multiple lines of different oscillator strengths. Further details are given in the description of the results for individual absorbers in Section~\ref{sec:result}.

We also note that the placement of the quasar continuum is difficult at these high redshifts due to the blending with the highly dense Ly-$\alpha$ forest lines. In many cases, at least one wing of the DLA line is blended with Ly-$\alpha$ forest lines. Moreover, the background quasars at these high redshifts are faint, making the spectra relatively noisy. Furthermore, the approach of doing a $\chi^2$ minimization to estimate the H I column density and its uncertainty gives unrealistically low uncertainties as discussed by \citet{Prochaska et al. 2003b}. As is standard in such situations, we constructed a series of profiles for H~I Ly-$\alpha$, Ly-$\beta$, and/or Ly-$\gamma$ lines for a range of H I column density values, all centered at the redshift of the metal absorption lines, and compared them with the observed data. The comparisons took into account, for each available Lyman series line, both the wings and the core (which is essential for constraining the H I column density when the wings are blended). The H I column densities thus derived are fairly robust, since at the resolution of X-shooter and ESI, much of the Ly-$\alpha$ forest is resolved. The uncertainties in the H I column densities were estimated by examining the range of values for which the fitted profiles are consistent with the observed data within the noise level. 

 No ionization corrections were applied to the abundances derived from the observed metal ions and H I. The primary reason for this is that these high-$N_{\rm H\, I}$ systems are not expected to be much ionized. Furthermore, the ionization potential of the main species used here for metallicity determination (O I) is slightly higher than that of H I; so charge-transfer reactions keep O I/ H I = O/H, thus making relative ionization effects minimal.

\subsection{Determination of Velocity Dispersion}
A strong correlation between metallicity and velocity dispersion has been reported for lower redshift DLAs and sub-DLAs \citep[e.g.][]{ Peroux et al. 2003, Ledoux et al. 2006, Moller et al. 2013, Som et al. 2015}. In fact, the metallicity, velocity dispersion, and redshift values are found to lie in a fundamental plane \citep[]{Neeleman et al. 2013}. Furthermore, the velocity-metallicity relations for DLAs and sub-DLAs appear to be distinct \citep[e.g.][]{Meiring et al. 2007, Meiring et al. 2009, Som et al. 2015, Quiret et al. 2016}. If the velocity dispersion is linked with mass in some way then the metallicity vs. velocity dispersion relation can be used to infer the mass-metallicity relation of absorber host galaxies. On the other hand, the velocity dispersion may arise in part in inflows, outflows or turbulent motion. In either case, it is helpful to test whether the high redshift velocity-metallicity relation follows the trend for lower redshift systems, or whether it has evolved. We have determined the velocity width $\Delta V_{90}$ which is the velocity range within which $90$ percent of the absorption is contained. The velocity dispersions were measured using unsaturated and relatively weak lines whenever possible, as described in \citet{Quiret et al. 2016} and \citet{Morrison et al. 2016}. We note, however, that the determination of $\Delta V_{90}$ is somewhat uncertain, because different ions can give different values. We describe this below in more detail in the discussion of the individual systems.
\begin{table*}
	\centering
	\caption{Summary of targets and observations}
	\label{tab:sum_obs}
	\begin{tabular}{cccccccc} 
		\hline
		QSO  & m$_i$ & z$_{em}$ & z$_{abs}$ & Instrument & \parbox[t]{0.5in}{Exposure  \par Time (s)\strut} & \parbox[t]{0.7in}{Wavelength \par Coverage({\AA})\strut} & \parbox[t]{0.8in}{Spectral\par Resolution (R)\strut}\\  
		\hline
		Q0231-0728  & 19.48 & 5.423 & 5.335 & Keck ESI & $2\times1800$ & 4000-10,000 & 5400\\
		Q0824+1302 & 19.95 & 5.212 & \parbox[t]{0.3in}{4.809 \par 4.829 \strut} & VLT X-Shooter & $3\times2400$ & 3000-25,000 & \parbox[t]{0.7in}{VIS: 8800\par NIR: 5300\strut}\\
		
		\hline
	\end{tabular}
\end{table*}

\subsection{Determination of Dust Depletion}
\citet{Jenkins 2009} developed a technique for estimating the extent of dust depletion using observations of multiple elements. Based on observations of 17 chemical elements (C, N, O, Mg, Si, S, P, Cl, Ti, Cr, Mn, Fe, Ni, Cu, Zn, Ge, and Kr) along the sightlines to 243 stars in the Milky Way, he established that depletion of elements on dust grains can be well-characterized in terms of three parameters. One of these parameters, $F_{*}$, represents a generalized depletion strength for the particular sightline being analyzed. The scale of $F_{*}$ is arbitrary, but it lies in the range of 0 to 1 for most sightlines in the Milky Way, although warm ionized gas in the Milky Way shows $F_{*} = -0.1$ \citep[]{Draine 2011}. The higher the value of $F_{*}$, the more severe is the depletion.  The other two parameters $A_{x}$ and $B_{x}$ are unique constants for each element X and were derived empirically by \citet{Jenkins 2009}.  $A_{x}$ is the propensity of the element X to have a higher depletion value as the line-of-sight depletion factor $F_{*}$ increases. In terms of these parameters and a zero-point offset $Z_{x}$ [added to make the measurement errors in $A_{x}$ and $B_{x}$ independent of each other, and also derived empirically by \citet{Jenkins 2009}], the gas-phase abundance of element X is described by the relation [X$_{gas}$/H] = $B_{x} + A_{x} (F_{*} - Z_{x})$. Thus, the coefficients $A_{x}$ and $B_{x}$ denote the slope and the offset in the linear relationship of depletion vs. $F_{*}$. In an extension of this work, \citet*{Jenkins 2017} presented the values of $A_{x}$, $B_{x}$, and $Z_{x}$ in the Small Magellanic Cloud (SMC) for 9 elements, and suggested that these values would be better suited for studies of depletion patterns in the low-metallicity DLA and sub-DLA absorbers in the distant universe.  

As further explained in Appendix A of \citet{Quiret et al. 2016}, the observed abundance of element X relative to the solar level can be expressed as [X/H]$_{\rm obs}$ - $B_{x}$ + $A_{x} z_{x}$ = [X/H]$_{\rm intrinsic}$ + $A_{x} F_{*}$.  In other words, in a plot of the quantity [X/H]$_{\rm obs}$ - $B_{x}$ + $A_{x} Z_{x}$ vs. $A_{x}$, the slope is the line-of-sight depletion factor $F_{*}$, and the intercept is [X/H]$_{\rm intrinsic}$, the intrinsic depletion-corrected metallicity derived from element X. In this paper, we adopt the values of $A_{x}$, $B_{x}$ and $Z_{x}$ for O and C from \citet{Jenkins 2009} and for Si and Fe from \citet*{Jenkins 2017} to estimate the $F_{*}$ value and the intrinsic metallicity for each absorber. Furthermore, we take the uncertainties in $A_{x}$ and $B_{x}$ into account, while estimating the uncertainties in $F_{*}$ and [X/H]$_{\rm intrinsic}$.
\section{Results for individual absorbers}
\label{sec:result}

\subsection{Absorber at $z=5.335$ toward Q0231-0728}
\begin{figure}
	
	\includegraphics[width=\columnwidth]{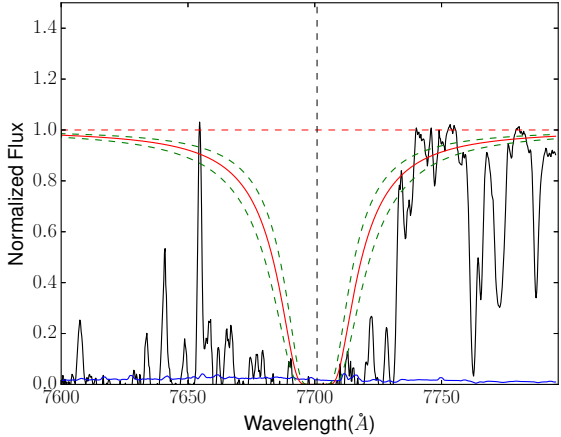}
    \caption{Voigt profile fitting for hydrogen Lyman$-\alpha$ for the $z=5.335$ absorber toward Q0231-0728. The observed normalized flux is shown in black and the fitted profile for log $N_{\rm H\, I}=20.10$ is shown in red. The green dashed lines above and below the fitted profile represent $\pm0.15$ dex uncertainty. The $1\sigma$ error in the normalized spectrum is shown in blue at the bottom. The vertical dashed line represents the centre of the profile and the horizontal dashed line in red shows the continuum level.}
    \label{fig:HI1216_5335}
\end{figure}

\begin{figure*}
\begin{tabular}{l}
\includegraphics[scale=0.45, angle=90]{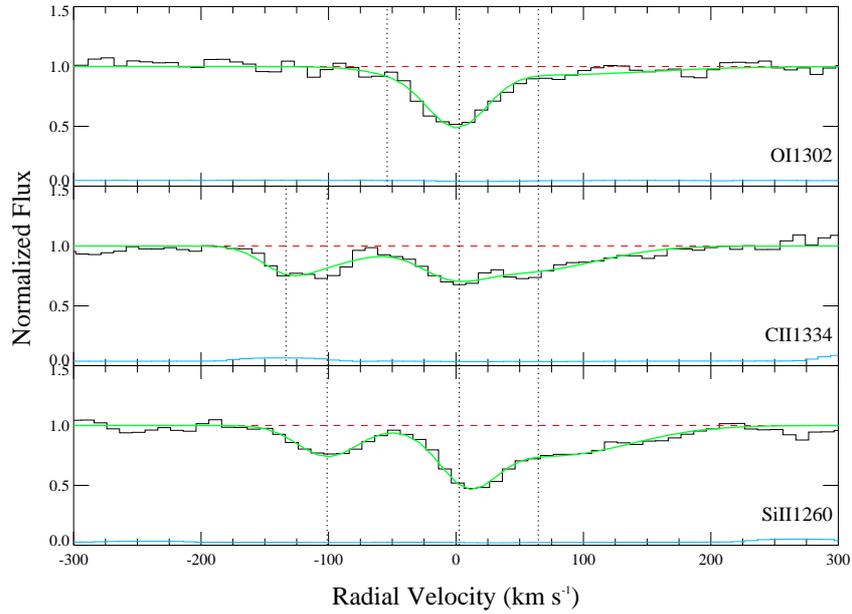}
\end{tabular}	

    \caption{ Velocity plots for metal lines in the $z=5.335$ absorber toward Q0231-0728. In each panel, the data are shown in black and the fitted profiles are shown in green. The blue line at the bottom show the $1\sigma$ error in the normalized flux and the horizontal line in red represents the continuum level. The vertical dotted lines show the different velocity components included in the profile fits.}
    \label{fig:metals_5335}
\end{figure*}

Voigt profile fitting to the Lyman$-\alpha$ line gave a column density log $N_{\rm H\, I} = 20.10\pm0.15$~(Fig. \ref{fig:HI1216_5335}). This absorber is thus close to the boundary between DLAs and sub-DLAs. Absorption features of ions O I, C II and Si II were detected in this system. The results from Voigt profile fitting~(Fig. \ref{fig:metals_5335}) for these metal lines are summarized in Table \ref{tab:voigt_5335}. Total element abundances  for O, C and Si ~(Table \ref{tab:metals_5335}) were calculated using O I $\lambda$1302, C II $\rm \lambda$1334 and Si II $\lambda$1260 respectively. We were not able to determine the abundance of Fe for this absorber, because all the Fe II lines either fall outside the wavelength coverage or are severely blended with the Lyman$-\alpha$ forest. 
\subsubsection{Metallicity and Relative Abundances}
The metallicity for this system based on the measurement of O is $-2.24\pm0.16$ dex. The intrinsic metallicity determined using the \citet{Jenkins 2009} prescription is $-2.28\pm0.17$ dex, which is consistent with the [O/H] value. The $F_{*}$ value that exhibits the extent of dust depletion was found to be $-0.27\pm0.24$. C is slightly underabundant compared to O with [C/O]= $-0.11\pm0.07$ dex. While [Si/C]($0.13\pm0.06$ dex) is enhanced, [Si/O]($0.02\pm0.06$ dex) is basically at the solar level, which is very different from values observed for the MW warm ISM, i.e., [Si/O] = -0.31 dex and [Si/C] = -0.22 dex \citep[]{Jenkins 2009}. Both Si and C have absorption features at $z=5.33286$ but O does not show any components at that redshift. Likewise, only C absorption is seen at $z=5.33218$, with no corresponding Si or O. This peculiarity in the abundance pattern might be an indicator of an unusual nucleosynthetic process at this higher redshift or may be the result of photoionization. 

\begin{table*}
	\centering
	\caption{Results of Voigt profile fitting for different elements in the $z=5.335$ absorber toward Q0231-0728}
	\label{tab:voigt_5335}
	\begin{tabular}{ccccccc} 
		\hline
		z  & b$_{\rm eff}$ (km s$^{-1})$  & log $N_{\rm O\, I}$ & log $N_{\rm C\, II}$  & log $N_{\rm Si\, II}$ \\
		\hline

	        $5.33218\pm0.00001$ & $10.20\pm2.12$ &   & $13.36\pm0.17$ &  \\	
	        $5.33286\pm0.00004$ & $18.37\pm4.73$ &   & $13.40\pm0.11$ & $12.65\pm0.03$ \\		$5.33385\pm0.00002$ & $12.71\pm2.36$ &  $13.24\pm0.31$ &   &  \\
		$5.33505\pm0.00003$ & $13.53\pm3.69$ &  $14.41\pm0.05$ & $13.54\pm0.10$ & $12.91\pm0.06$\\		
		$5.33636\pm0.00015$ & $72.73\pm7.14$ &  $13.89\pm0.15$ & $13.83\pm0.07$ & $13.08\pm0.05$ \\	
	\hline
		Total log N &  & $14.55\pm0.05$& $14.18\pm0.05$ & $13.39\pm0.03$ \\
		\hline	
		\end{tabular}
\end{table*}

\begin{table*}
	\centering
	\caption{Abundances of different elements in the $z=5.335$ absorber toward Q0231-0728}
	\label{tab:metals_5335}
	\begin{tabular}{ccc} 
		\hline
		Element & [X/H] & [X/O]\\
		\hline
		O & $-2.24\pm0.16$ &  \\
		C & $-2.35\pm0.16$ & $-0.11\pm0.07$ \\
		Si & $-2.22\pm0.15$ & $0.02\pm0.06$\\
		\hline
	\end{tabular}
\end{table*}
\subsubsection{Gas Kinematics}

\begin{figure*}
\begin{tabular}{lll}
\includegraphics[scale=0.32, angle=90]{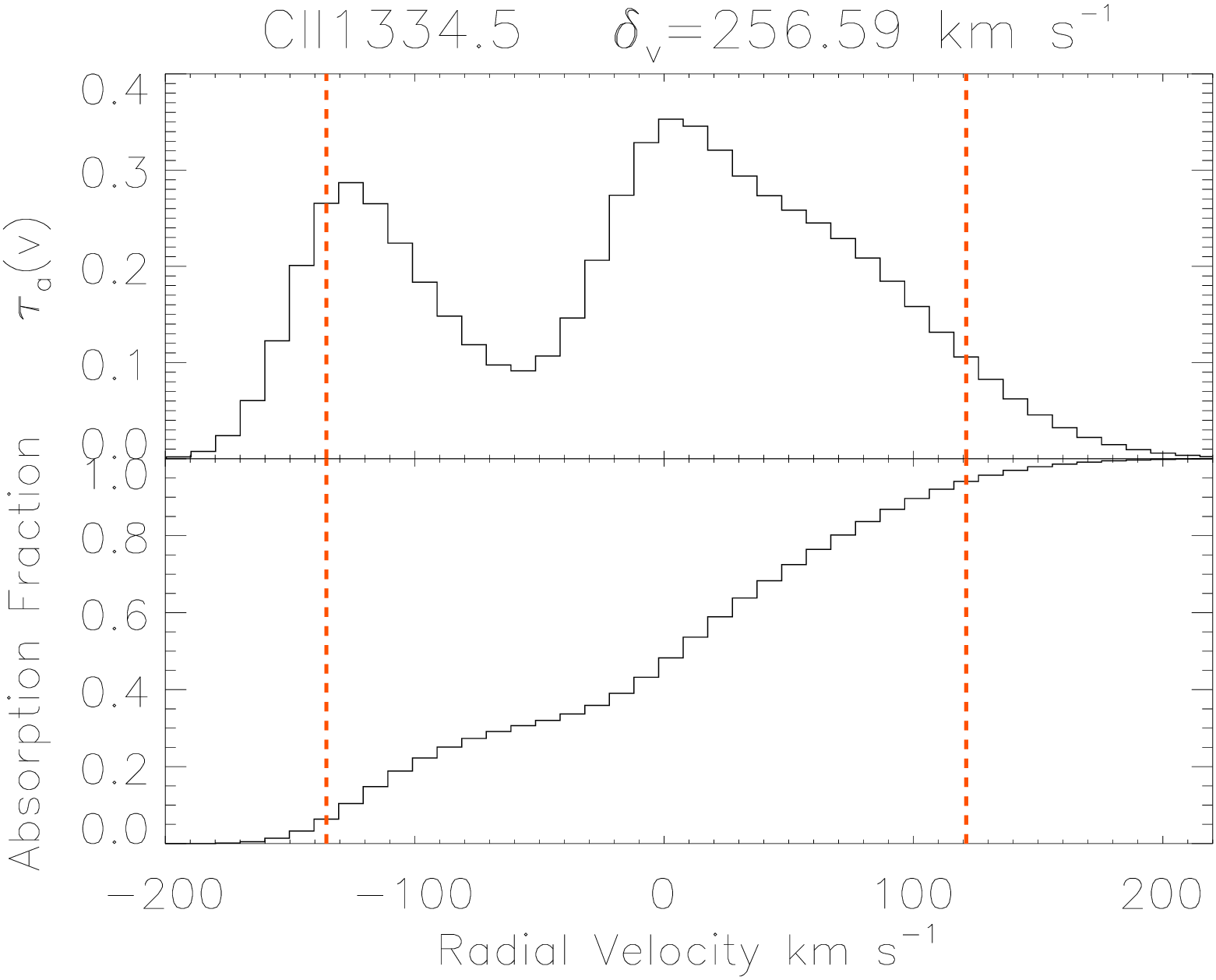}
\includegraphics[scale=0.32, angle=90]{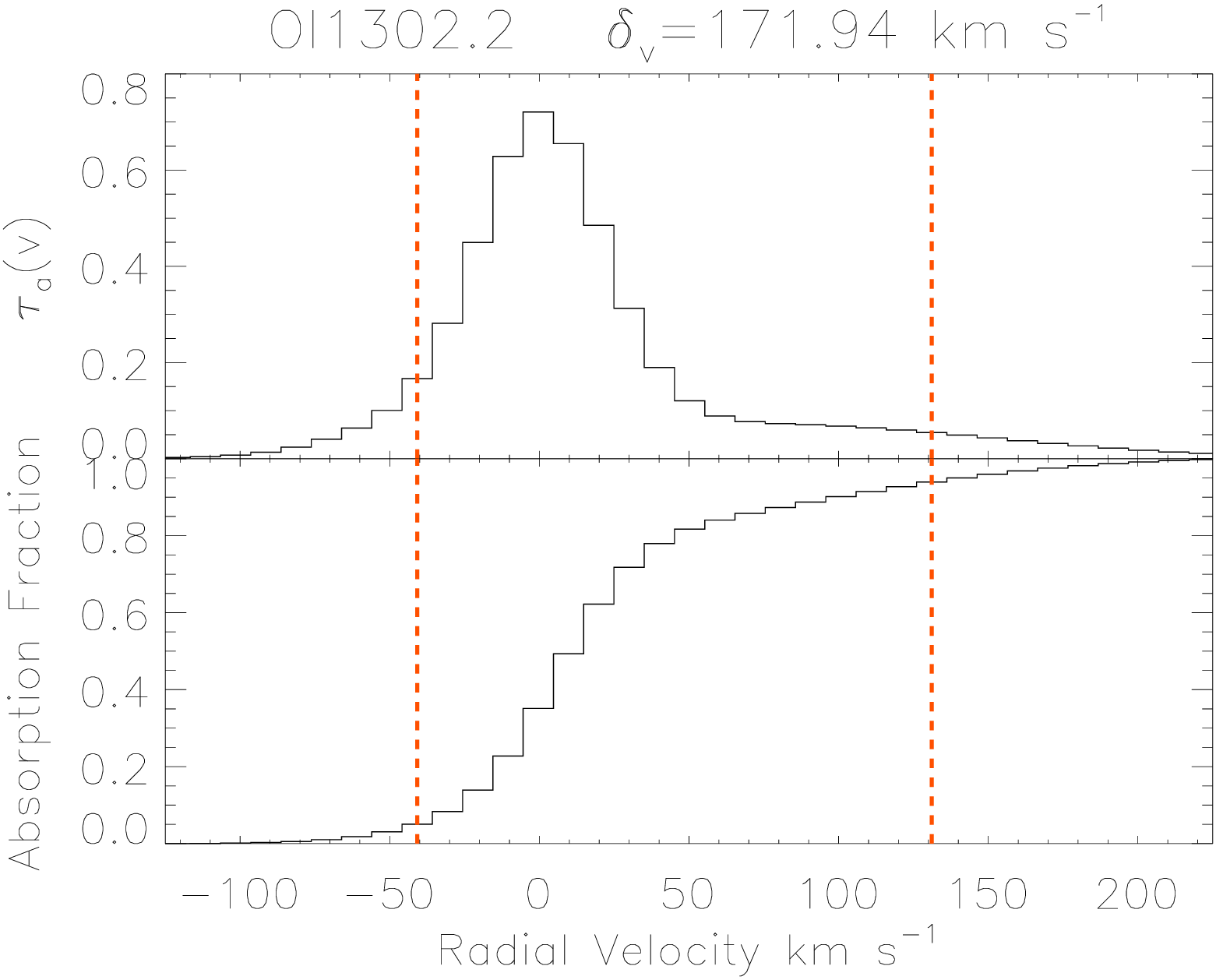}\\[2\tabcolsep]
\includegraphics[scale=0.35, angle=90]{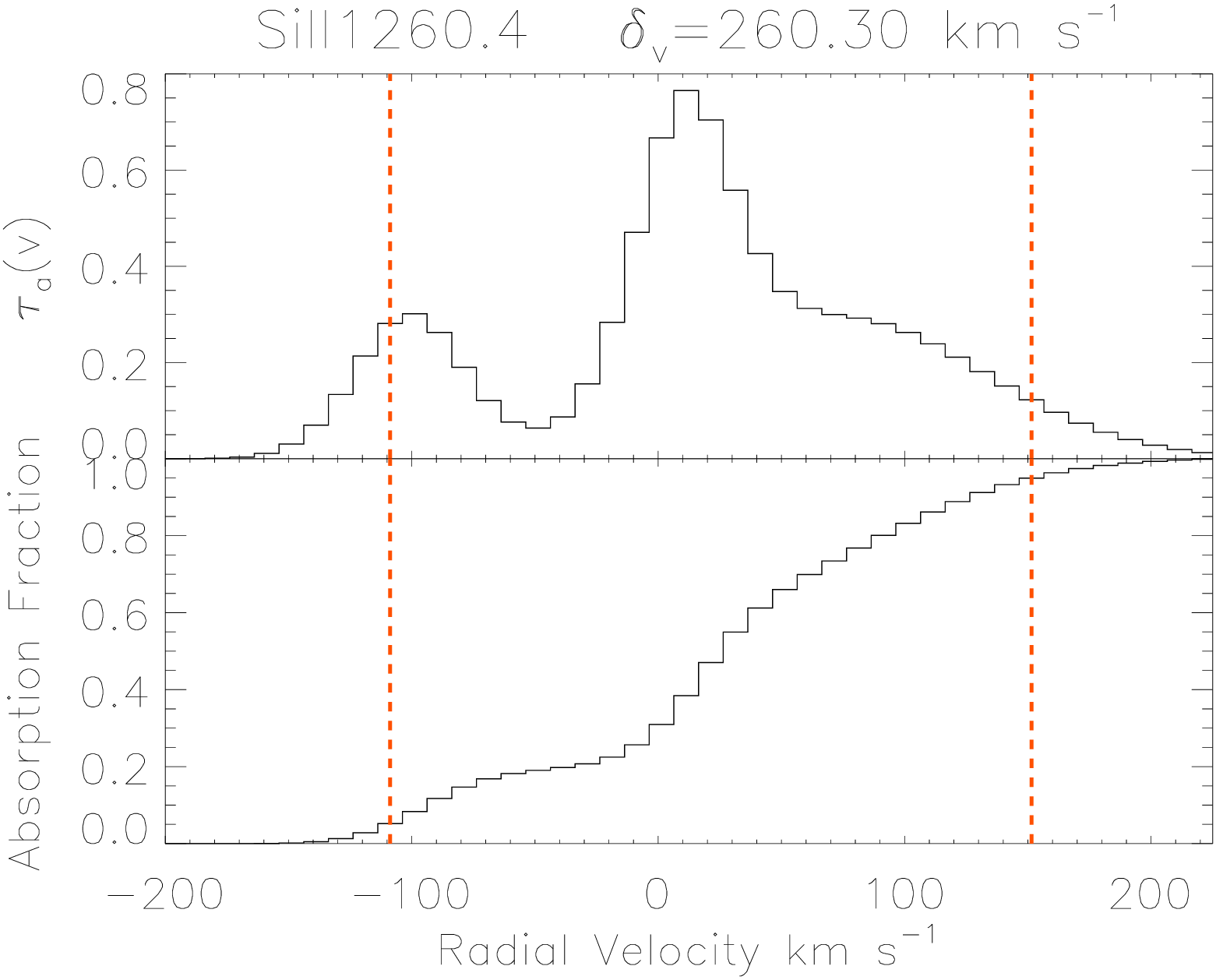}
\end{tabular}
\caption{Determination of velocity dispersion for C II $\lambda$1334, O I $\lambda$1302  and Si II $\lambda$1260 in the $z=5.335$ absorber toward Q0231-0728 based on our Voigt profile fits. In each case, the upper panel shows the optical depth profile and the lower panel shows the absorption fraction as a function of velocity. The 5 percent and 95 percent levels of the absorption are marked with two vertical dotted lines in red.}
\label{fig:5335_kinematics}
\end{figure*}

 Fig. \ref{fig:5335_kinematics} shows the optical depth and the absorption fraction as a function of radial velocity for the fitted profiles of Si II $\lambda$1260, C II $\lambda$1334 and O I $\lambda$1302. O I absorption for this absorber is limited to a narrower velocity range of 172 km s$^{-1}$ than C II absorption and SiII absorption which have similar velocity dispersion values (257 km s$^{-1}$ and 260 km s$^{-1}$, respectively). This situation is similar to the smaller O~I dispersion range than C II and Si II reported in a $z = 4.977$ absorber \citep[]{Morrison et al. 2016}. Here we adopt the velocity dispersion measured using Si II $\lambda$1260. \\
 
A correlation between velocity dispersion and metallicity has been observed \citep[e.g.][]{Ledoux et al. 2006, Meiring et al. 2007, Som et al. 2015, Quiret et al. 2016}.  Using the metallicity-velocity relation [X/H] = $(1.52\pm0.08)$ log$\Delta V_{90} - (4.20\pm0.16)$ observed for lower redshift DLAs \citep[]{Quiret et al. 2016}, an absorber with $\Delta V_{90}$ = 260 km s$^{-1}$ would typically have a metallicity of -0.53 dex. The metallicity-velocity relation [X/H] = $(1.61\pm0.22)$ log$\Delta V_{90} - (3.94\pm0.45)$ observed for sub-DLA would imply metallicity of -0.05 dex. Thus, the observed metallicity of the $z=5.335$ absorber is far below the typical metallicity indicated by the metallicity vs. velocity dispersion relations for both DLAs and sub-DLAs at lower redshifts. Despite being more metal poor, this absorber has a significantly larger velocity dispersion than observed for the absorber at $z=4.977$ \citep[]{Morrison et al. 2016}, which has a metallicity of -2.05 dex and a velocity dispersion of 84 km s$^{-1}$ for Si II.

\subsection{Absorber at $z=4.809$ toward Q0824+1302}

\begin{figure*}
\begin{tabular}{ll}
\includegraphics[scale=0.4]{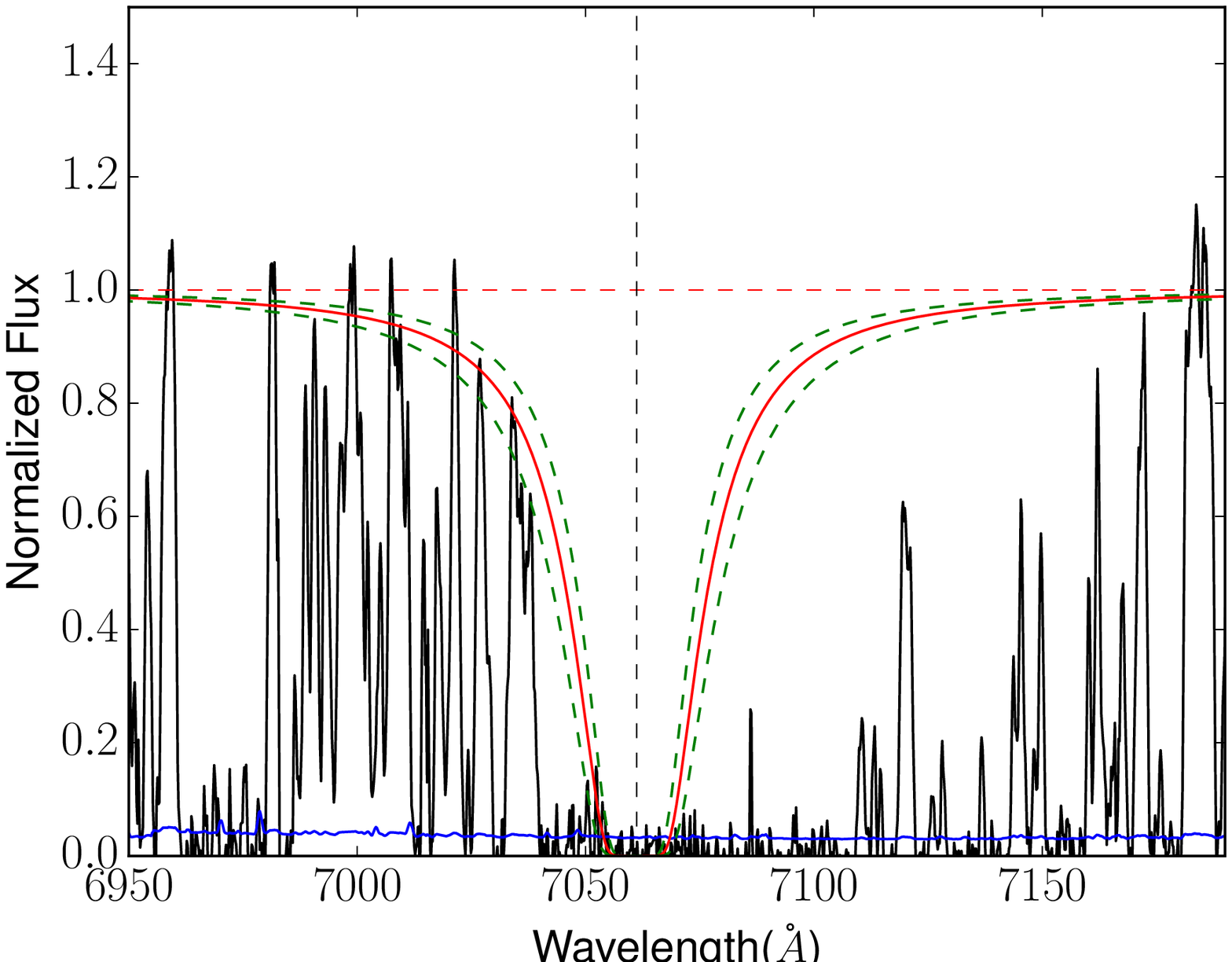}
&
\includegraphics[scale=0.4]{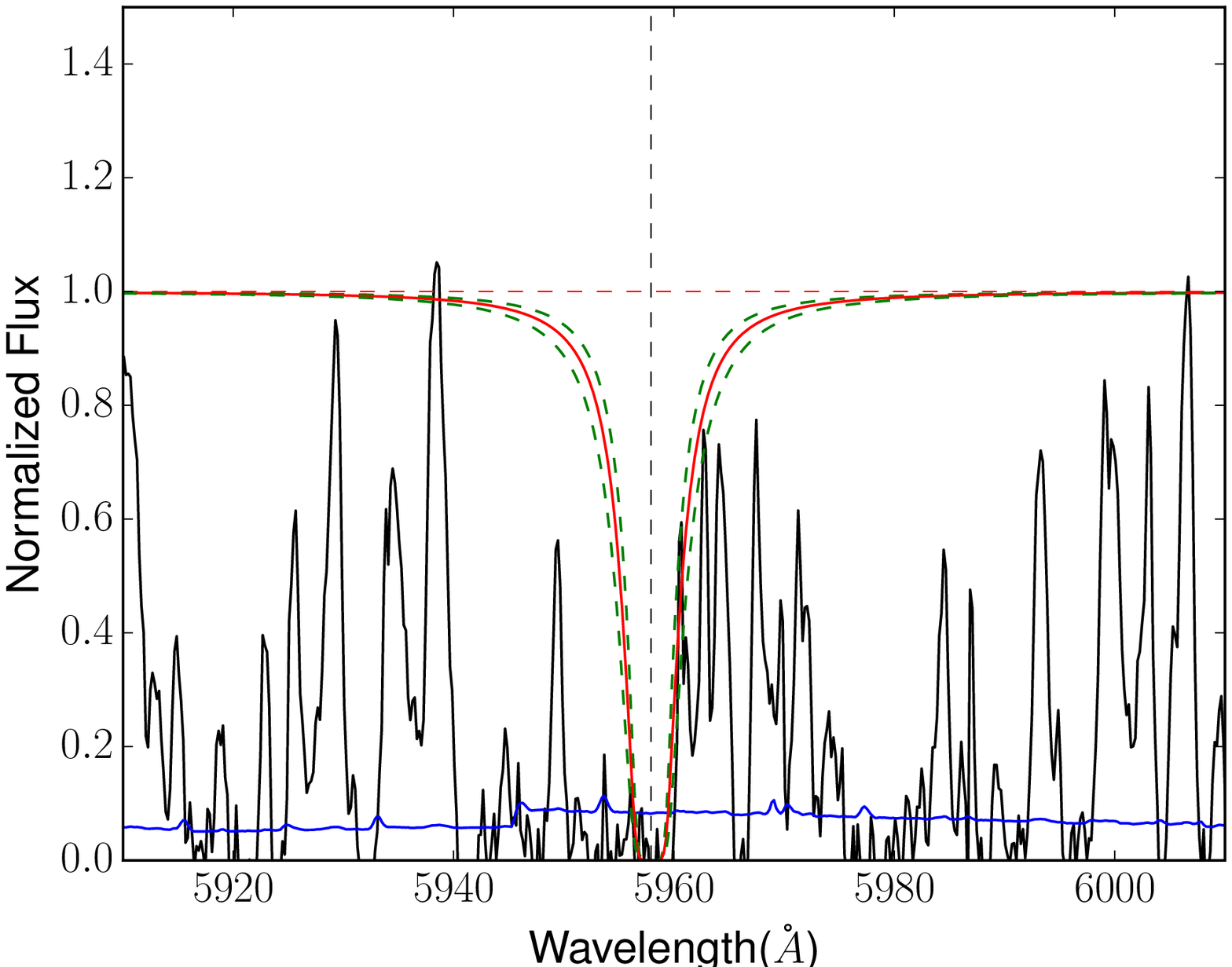}
\end{tabular}
\caption{Voigt profile fitting for hydrogen Lyman$-\alpha$ (Left) and hydrogen Lyman$-\beta$ (Right) for the $z=4.809$ absorber toward Q0824+1302. Observed normalized flux is shown in black and the fitted profile for log $N_{\rm H\, I}=20.10$ is shown in red. Two green dashed lines above and below the fitted profile represent $\pm0.15$ dex uncertainty. The $1\sigma$ error in the normalized spectrum is shown in blue at the bottom. In each panel, the vertical dashed line represents the centre of the profile, and the horizontal dashed line in red shows the continuum level.}
\label{fig:4809_lyman}
\end{figure*}

\begin{figure*}
\begin{tabular}{l}
\includegraphics[scale=0.6, angle=90]{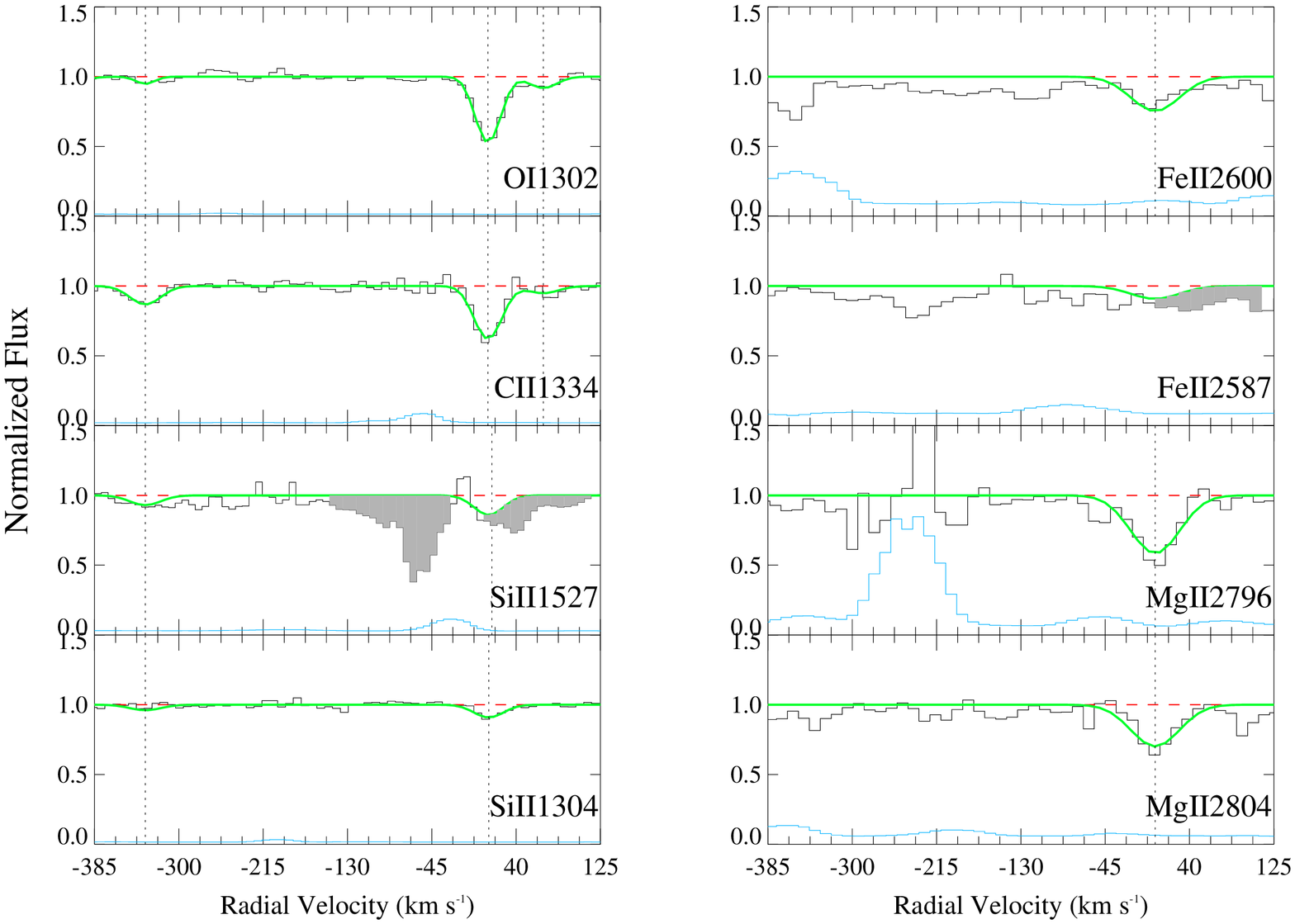}
\end{tabular}
\caption{Velocity plots for metal lines for the absorber at $z=4.809$ toward Q0824+1302. In each panel, the data are shown in black and the fitted profiles are shown in green. The blue line at the bottom shows the $1\sigma$ error in normalized flux and the horizontal line in red represents the continuum level. Unrelated absorption features are shaded in grey. In each panel, the vertical dotted lines show the different velocity components included in the profile fits.}
\label{fig:4809_metals}
\end{figure*}

Voigt profile fits to the Lyman-$\alpha$ and Lyman-$\beta$ lines for this absorber gave an H I column density of log $N_{\rm H\, I} = 20.10 \pm 0.15$  and are shown in Fig. \ref{fig:4809_lyman}. This absorber is thus also close to the boundary between DLAs and sub-DLAs. Absorption features of O I, C II, Si II and Fe II were detected in this system. The results from Voigt profile fitting~(Fig. \ref{fig:4809_metals}) for these metal lines are summarized in Table~\ref{tab:voigt_4809}. Table \ref{tab:metals_4809} summarizes the total element abundances for O, C, Si and Fe. The Fe abundance was calculated fitting  Fe II $\lambda$$\lambda$ 2587, 2600 together, while the Si abundance was determined by fitting Si II $\lambda$1304 and Si II $\lambda$1527 together. 
\subsubsection{Relative Abundances}
The metallicity for this system based on the measurement of O is $-2.51\pm0.16$ dex. The intrinsic metallicity determined using the prescription of \citet{Jenkins 2009} is $-2.71\pm0.17$ dex, which is consistent with the [O/H] value. The $F_{*}$ value that measures the depletion strength is $-0.77 \pm 0.21$. C is underabundant compared to O with [C/O] = $-0.24\pm0.10$ dex, while [Fe/O] = $0.03\pm0.18$ dex is basically at the solar level. The [C/O] for this absorber is similar to the value obtained for the sub-DLA absorber at $z=4.977$ with [C/O] = $-0.25\pm0.09$ dex \citep{Morrison et al. 2016}. However, the [Fe/O] ratio is much higher than that in the latter absorber ([Fe/O] = $-0.58\pm0.14$ dex, \citet{Morrison et al. 2016}). Si is enriched relative to O and C with [Si/O] = $0.20\pm0.15$ dex and [Si/C] = $0.70\pm0.16$ dex respectively. Also, the abundance of Si relative to Fe is consistent with being solar ([Si/Fe] = $0.17\pm0.22$ dex). The overall enhancement of Si with respect to C and O makes this absorber very peculiar suggesting the effect of an unusual nucleosynthetic history.

\begin{table*}
	\centering
	\caption{Results of Voigt profile fitting for different elements in the $z=4.809$ absorber towards Q0824+1302}
	\label{tab:voigt_4809}
	\begin{tabular}{ccccccc} 
		\hline
		z  & b$_{\rm eff}$ (km s$^{-1}$)  & log $N_{\rm O\, I}$ & log $N_{\rm C\, II}$  & log $N_{\rm Si\, II}$  & log $N_{\rm Fe\, II}$  & log $N_{\rm Mg\, II}$  \\
		\hline

		$4.80082\pm0.00037$ & $26.37\pm5.64$ & $13.22\pm0.28$ &  &  &  & \\
		$4.80253\pm0.00013$ & $10.12\pm2.23$ & $13.01\pm0.29$ & $13.13\pm0.19$ & $12.74\pm0.28$\\
		$4.80900\pm0.00005$ & $10.23\pm1.25$ &  &  &  & $13.12\pm0.17$ & $13.11\pm0.09$\\
		$4.80922\pm0.00003$ & $14.36\pm2.30$ &  $14.18\pm0.06$ & $13.62\pm0.10$ & $13.16\pm0.16$\\	
                $4.81030\pm0.00019$ & $17.13\pm3.41$ & $13.13\pm0.29$ & $12.68\pm0.32$ &  & \\
	\hline
	Total log N &  & $14.28\pm0.06$& $13.77\pm0.09$ & $13.30\pm0.14$ & $13.12\pm0.17$ & $13.11\pm0.09$\\
	\hline
	\end{tabular}
\end{table*}

\begin{table*}
	\centering
	\caption{Abundances of different elements in the $z=4.809$ absorber toward Q0824+1302}
	\label{tab:metals_4809}
	\begin{tabular}{ccc} 
		\hline
		Element & [X/H] & [X/O]\\
		\hline
		O & $-2.51\pm0.16$ &  \\
		C & $-2.75\pm0.17$  & $-0.24\pm0.10$\\
		Si & $-2.31\pm0.20$ & $0.20\pm0.15$\\
		Fe & $-2.48\pm0.23$ & $0.03\pm0.18$\\
		\hline
	\end{tabular}
\end{table*}

\subsubsection{Gas Kinematics}

\begin{figure*}
\begin{tabular}{ll}
\includegraphics[scale=0.3, angle=90]{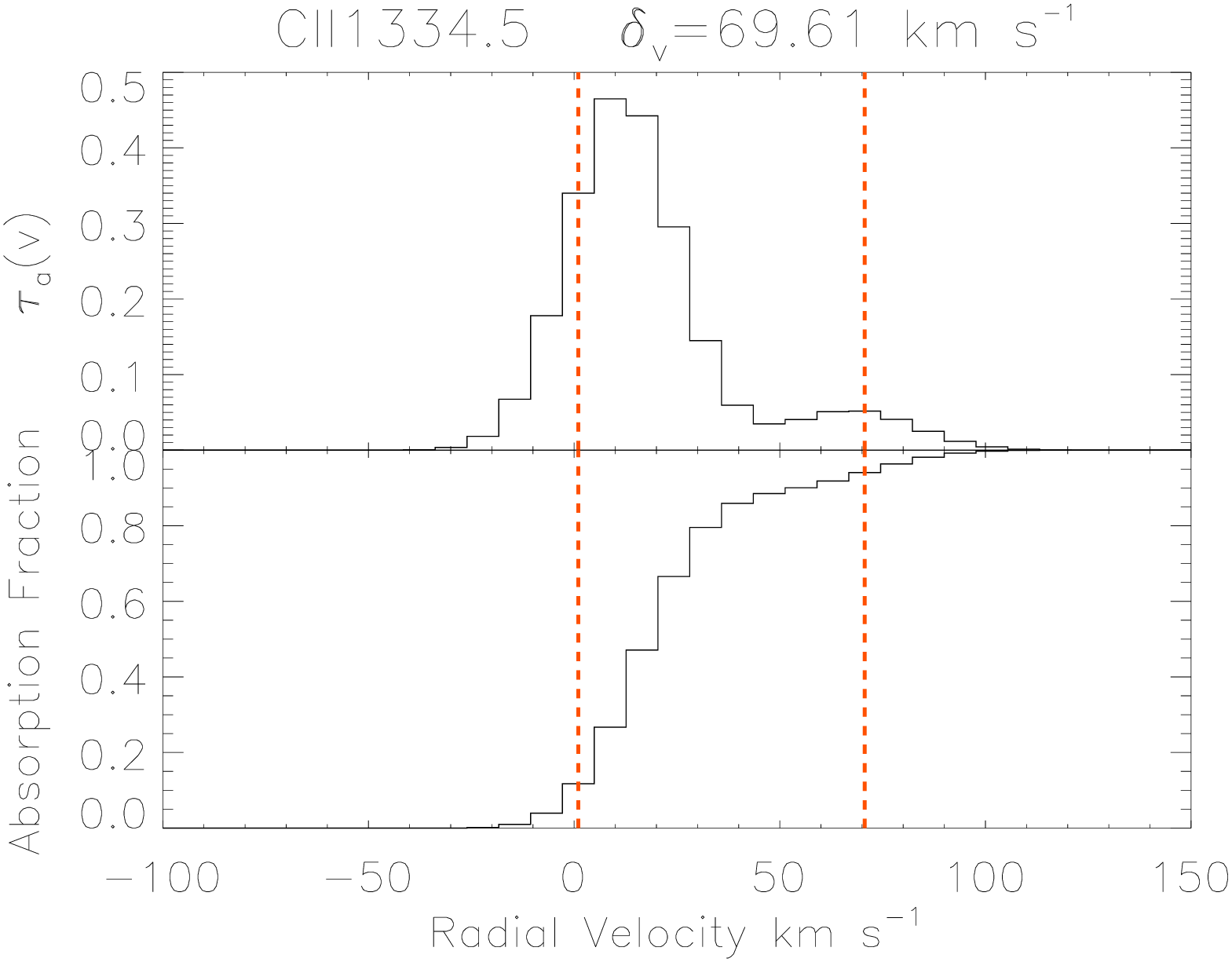}
&
\includegraphics[scale=0.3, angle=90]{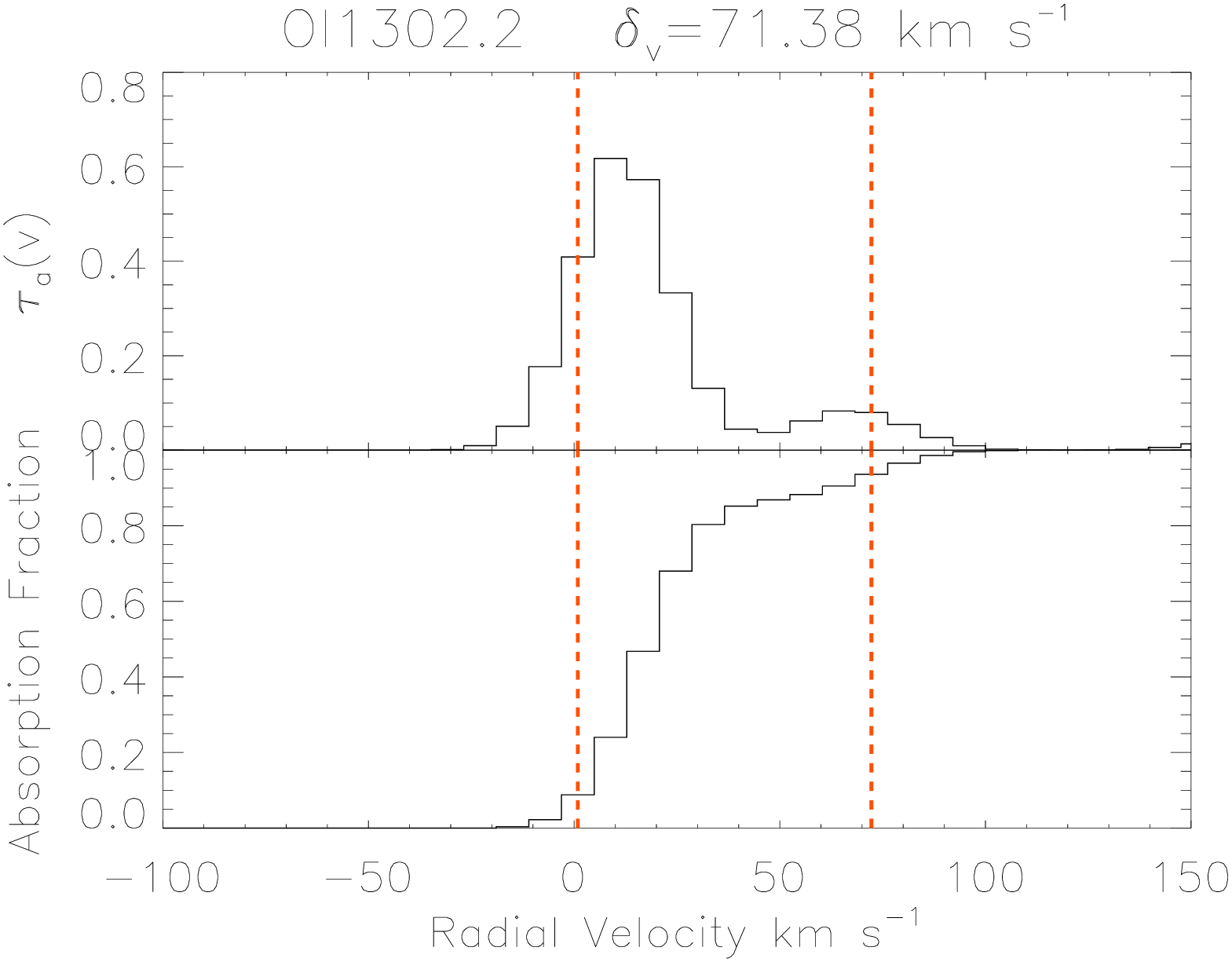}
\end{tabular}
\caption{Velocity dispersion for C II $\lambda$1334 and O I $\lambda$1302 lines in the $z=4.809$ absorber toward Q0824+1302 based on our Voigt profile fits. Upper panel shows the optical depth profile and lower panel shows the absorption fraction as a function of radial velocity. The 5 percent and 95 percent level of the absorption are marked with two vertical dotted lines in red. }
\label{Fig:4809_kinematics}
\end{figure*}

The C II absorption in this absorber is spread over a velocity width of 69 km s$^{-1}$~(Fig. \ref{Fig:4809_kinematics}).  An absorption feature can also be seen at a velocity of -344 km s$^{-1}$ but was excluded because of its small contribution to the overall absorption. The higher velocity components including the one at 75 km s$^{-1}$ may signify outflows. Based on the metallicity-velocity relation [X/H] = $(1.52\pm0.08)$ log $\Delta V_{90} - (4.20\pm0.16)$ observed for lower redshift DLAs \citep[]{Quiret et al. 2016}, an absorber with $\Delta V_{90}$ = 69 km s$^{-1}$ would typically have a metallicity of -1.39 dex. From the metallicity-velocity relation [X/H] = $(1.61\pm0.22)$ log $\Delta V_{90} - (3.94\pm0.45)$ observed for sub-DLAs, the predicted metallicity for a sub-DLA with $\Delta V_{90}$ = 69 km s$^{-1}$ would be typically -0.97 dex. The observed metallicity for this absorber is thus far below the typical values indicated by the metallicity vs. velocity dispersion relations for both DLAs and sub-DLAs at lower redshifts. \citet{Morrison et al. 2016} found a similar velocity width of 84 km s$^{-1}$ for a sub-DLA at $z=4.977$ that has a metallicity of -2.05 dex.

\subsection{Absorber at $z=4.829$ toward Q0824+1302}

\begin{figure*}
\begin{tabular}{ll}
\includegraphics[scale=0.4]{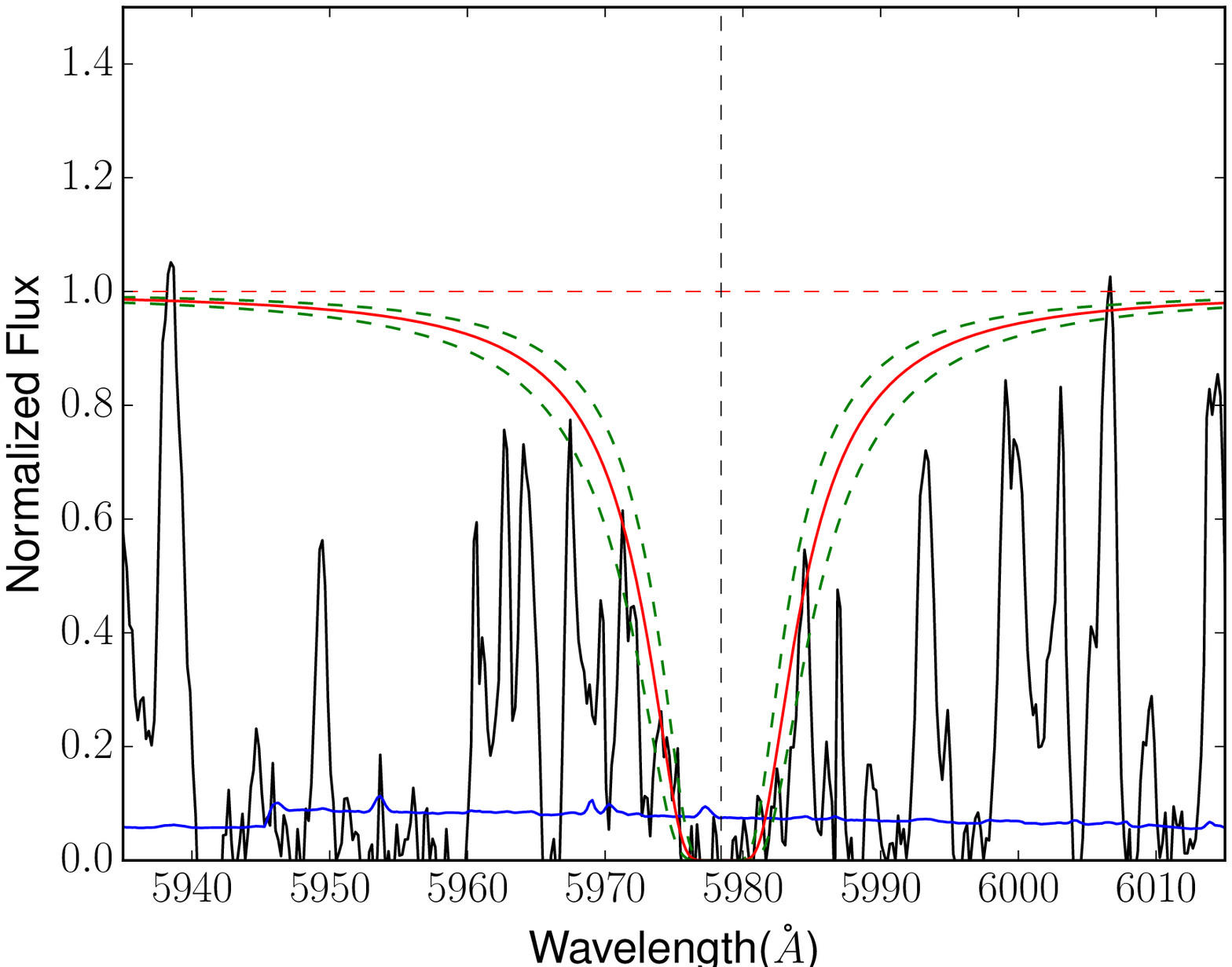}
&
\includegraphics[scale=0.4]{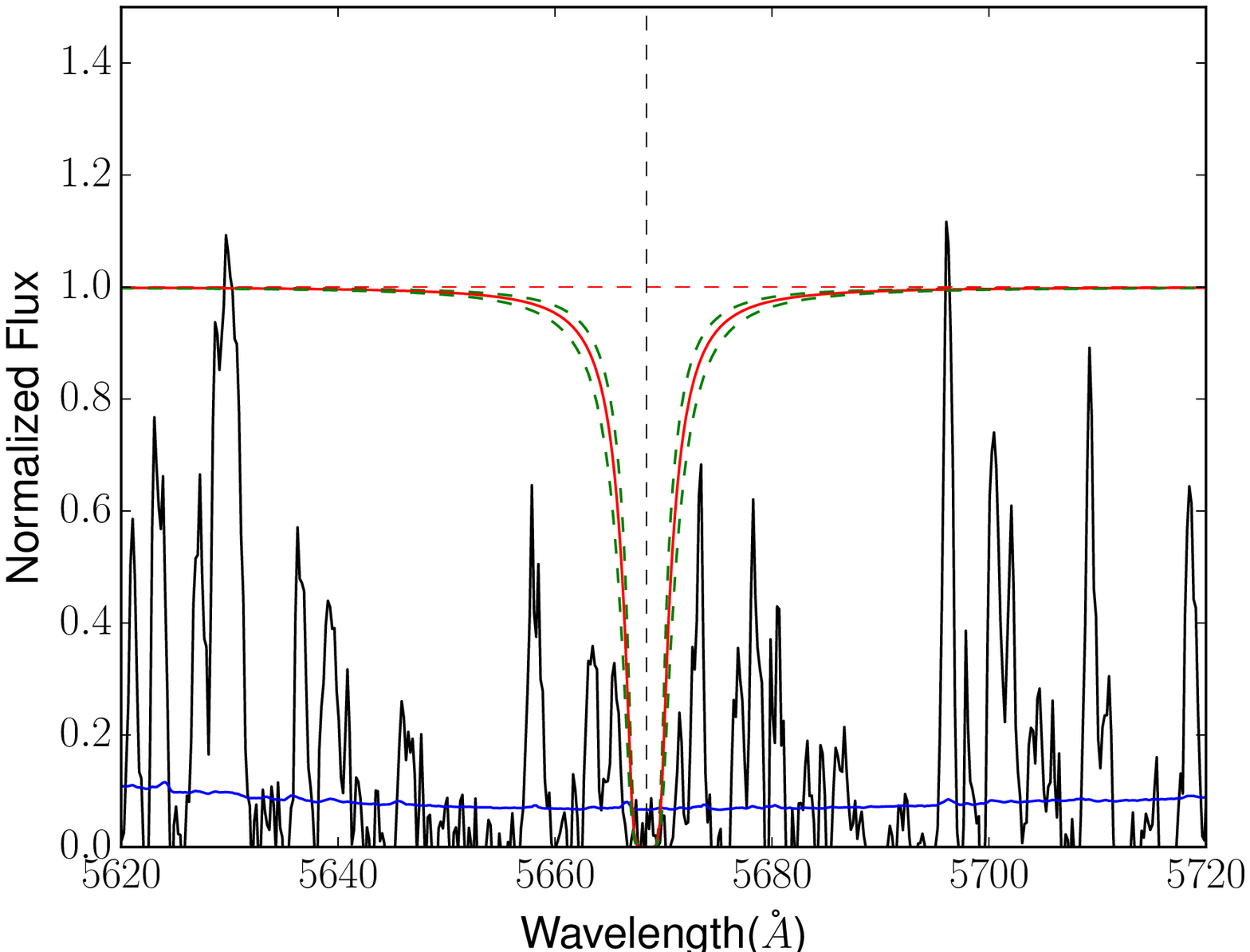}
\end{tabular}
\caption{Voigt profile fitting for hydrogen Lyman$-\beta$ (Left)  and hydrogen Lyman$-\gamma$ (Right) for the $z=4.829$ absorber toward Q0824+1302. Observed normalized flux is shown in black and the fitted profile for log $N_{\rm H\, I} = 20.80$ is shown in red. Two green dashed lines above and below the fitted profile represent $\pm0.15$ dex uncertainty. The $1\sigma$ error in the normalized spectrum is shown in blue at the bottom. In each panel, the vertical dashed line represents the centre of the profile, and the horizontal dashed line in red shows the continuum level.}
\label{fig:4829_lyman}
\end{figure*}

\begin{figure*}
\begin{tabular}{l}
\includegraphics[scale=0.5, angle=90]{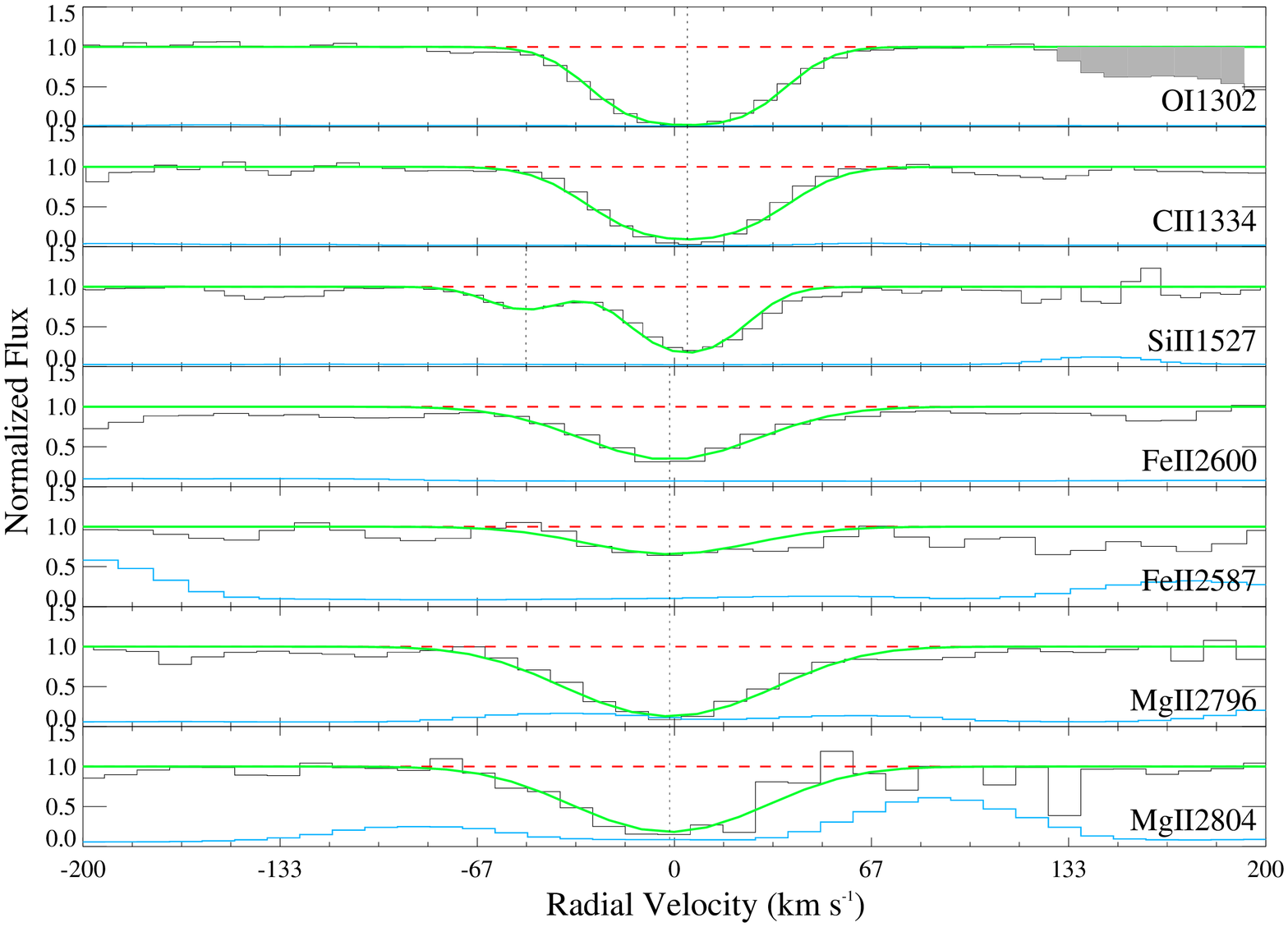}
\end{tabular}
\caption{ Velocity plots for metal lines for the absorber at $z=4.829$ toward Q0824+1302. In each panel, the data are shown in black and the fitted profiles are shown in green. The blue line at the bottom of each panel shows the $1\sigma$ error in the normalized flux and the horizontal line in red represents the continuum level. Unrelated absorption features are shaded in grey. The vertical dotted lines show the different velocity components included in the profile fits.}
    \label{fig:metals_4829}
\end{figure*}

\begin{figure*}
	
	\includegraphics[width=\columnwidth]{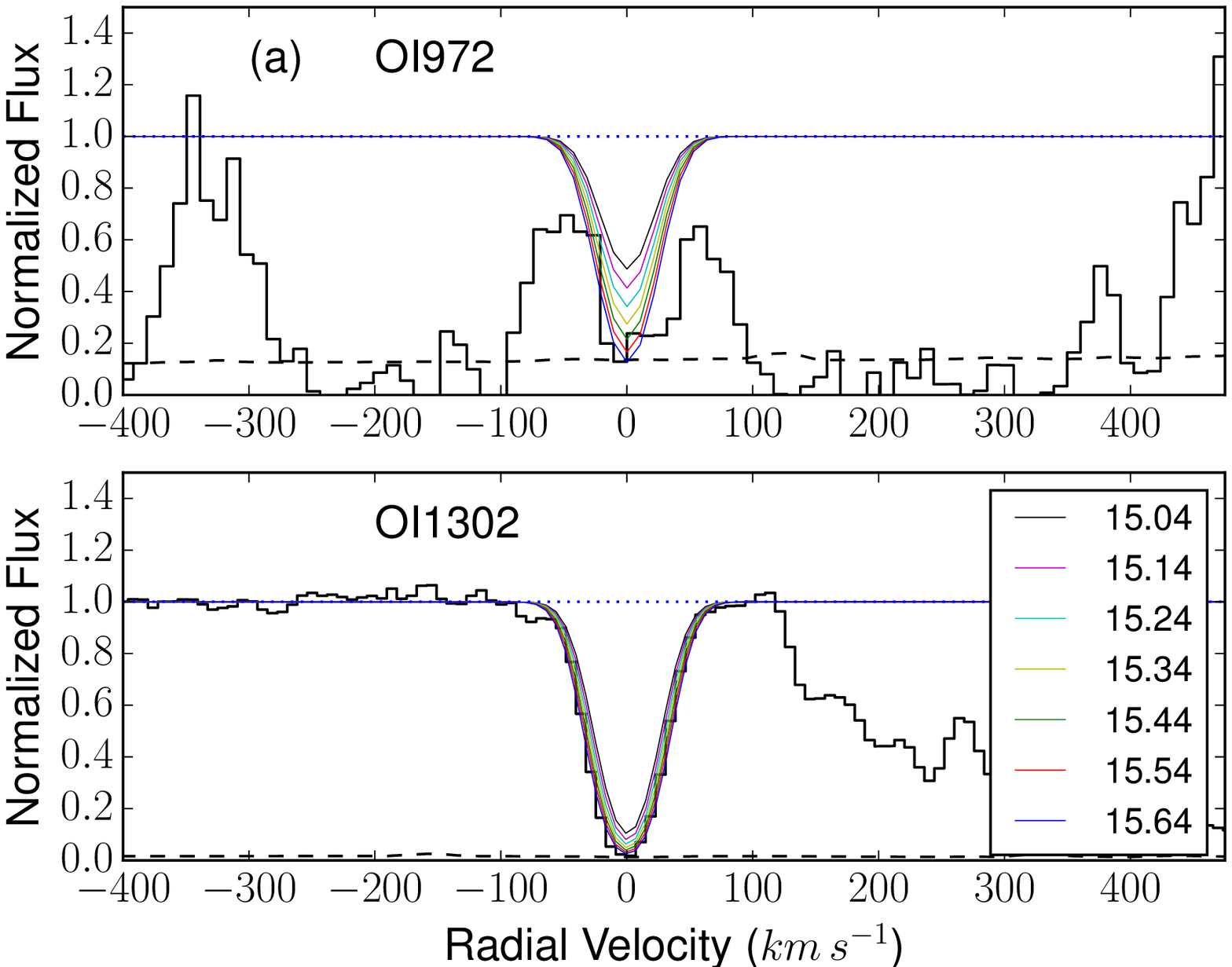}
	\includegraphics[width=\columnwidth]{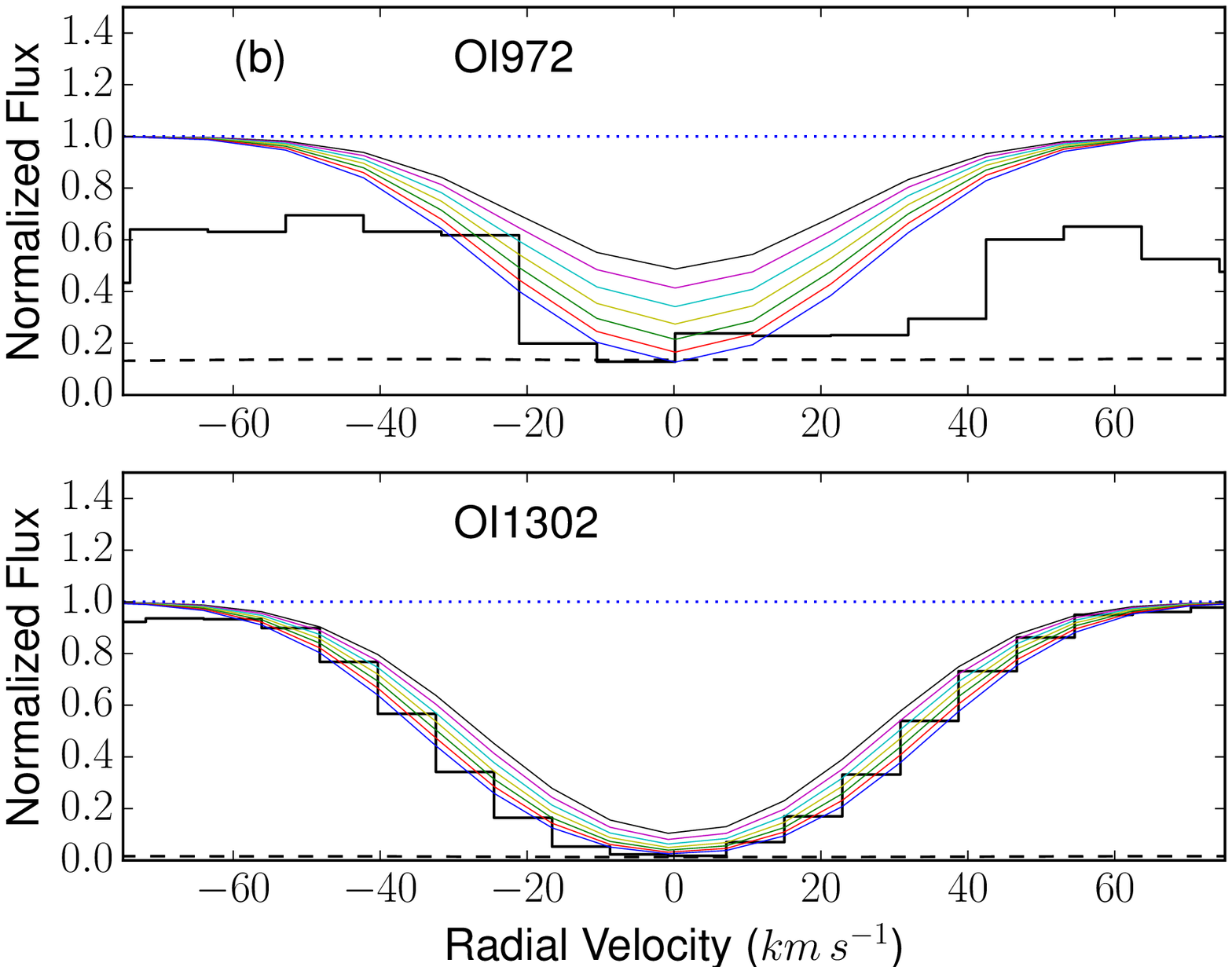}
	    \caption{ a) Velocity plots for O I $\lambda$$\lambda$ 972, 1302 metal lines showing the expected absorption profiles for different column density values with increment of 0.1 dex in each interval for the absorber at $z=4.829$ toward Q0824+1302. In each panel, the data are shown in black and the fitted profiles are shown in colours. The black dashed line at the bottom of each panel shows the $1\sigma$ error in normalized flux and the horizontal line in blue represents the continuum level. b) Same as (a) but showing a narrower velocity range for more clarity.}
    \label{fig:overplot}
\end{figure*}

Voigt profile fits to the Lyman-$\beta$ and Lyman-$\gamma$ lines for this absorber gave an H I column density of log $N_{\rm H\, I} = 20.80\pm0.15$ and are shown in Fig. \ref{fig:4829_lyman}. This absorber is thus a DLA. Absorption features of O I, C II, Si II and Fe II were detected in this system. The results from Voigt profile fitting~(Fig. \ref{fig:metals_4829}) for these metal lines are summarized in Table \ref{tab:voigt_4829}. Table~\ref{tab:metals_4829} summarizes the total element abundances  for O, C, Si and Fe. The Fe abundance was calculated fitting the Fe II $\lambda$$\lambda$ 2587, 2600 lines together. The O I column density obtained from O I $\lambda$1302 alone is log N$_{\rm O\, I} = 15.54\pm0.20$. To further improve the constraints on the O I column density, we use both O I $\lambda$972 and O I $\lambda$1302~(Fig. \ref{fig:overplot} a, b). O~I $\lambda$972 for this absorber falls in the Lyman-$\alpha$ forest, but is less saturated than O I $\lambda$1302 (which falls outside the Ly-$\alpha$ forest). We use both the cores and the wings of the lines to estimate that log $N_{\rm O\, I}$ cannot be much lower than 15.29 or much higher than 15.59. We adopt log $N_{\rm O\, I} = 15.44 \pm 0.15$. 
\begin{table*}
	\centering
	\caption{Results of Voigt profile fitting for different elements in the $z=4.829$ absorber towards Q0824+1302}
	\label{tab:voigt_4829}
	\begin{tabular}{ccccccc} 
		\hline
		z  & b$_{\rm eff}$ (km s$^{-1}$)  & log $N_{\rm O\, I}$ & log $N_{\rm C\, II}$  & log $N_{\rm Si\, II}$  & log $N_{\rm Fe\, II}$  & log $N_{\rm Mg\, II}$  \\
		\hline
		
		$4.82802\pm0.00012$ & $15.15\pm3.41$ &  & &$13.22\pm0.19$ \\
		$4.82908\pm0.00002$ & $28.98\pm1.63$ &  $15.44\pm0.15$ & $14.88\pm0.12$ & $14.17\pm0.09$ \\		
		$4.82891\pm0.00003$ & $32.35\pm2.09$ & & & & $13.82\pm0.13$ & $14.14\pm0.12$\\
		\hline
Total log N & & $15.44\pm0.15$ & $14.88\pm0.12$ & $14.22\pm0.08$ & $13.82\pm0.13$ &$14.14\pm0.12$\\
                \hline

	\end{tabular}
\end{table*}

\begin{table*}
	\centering
	\caption{Abundances of different elements in the $z=4.829$ absorber toward Q0824+1302}
	\label{tab:metals_4829}
	\begin{tabular}{ccc} 
		\hline
		Element & [X/H] & [X/O]\\
		\hline
		O & $-2.05\pm0.21$ &\\
		C & $-2.35\pm0.19$  & $-0.30\pm0.19$\\
		Si & $-2.09\pm0.17$ & $-0.04\pm0.17$\\
		Fe &$-2.48\pm0.20$ & $-0.43\pm0.20$\\
		\hline
	\end{tabular}
\end{table*}

\subsubsection{Relative abundances}
The metallicity for this system based on the measurement of O is $-2.05\pm0.21$ dex. The intrinsic metallicity determined using the \citet{Jenkins 2009} prescription is $-2.22\pm0.20$ dex which is consistent with the [O/H] value. The $F_{*}$ value that exhibits the extent of dust depletion was found to be $-0.45\pm0.22$. 
C and Fe are underabundant compared to O with [C/O] = $-0.30\pm0.19$ dex and [Fe/O] = $-0.43\pm0.20$ dex. The [C/O] ratio for this absorber is higher and the [Fe/O] ratio is similar compared to the $z = 4.977$ absorber from \citet[]{Morrison et al. 2016}, which has [C/O] = $-0.25\pm0.09$ and [Fe/O] = $-0.58\pm0.14$). The abundance of Si relative to O in the $z=4.829$ absorber is nearly solar, with [Si/O] = $-0.04 \pm 0.17$ dex; but Si is overabundant in this absorber relative to Fe, with [Si/Fe] = $0.39 \pm 0.15$ dex.  
\subsubsection{Gas Kinematics}
\begin{figure*}
\begin{tabular}{l}
\includegraphics[scale=0.35, angle=90]{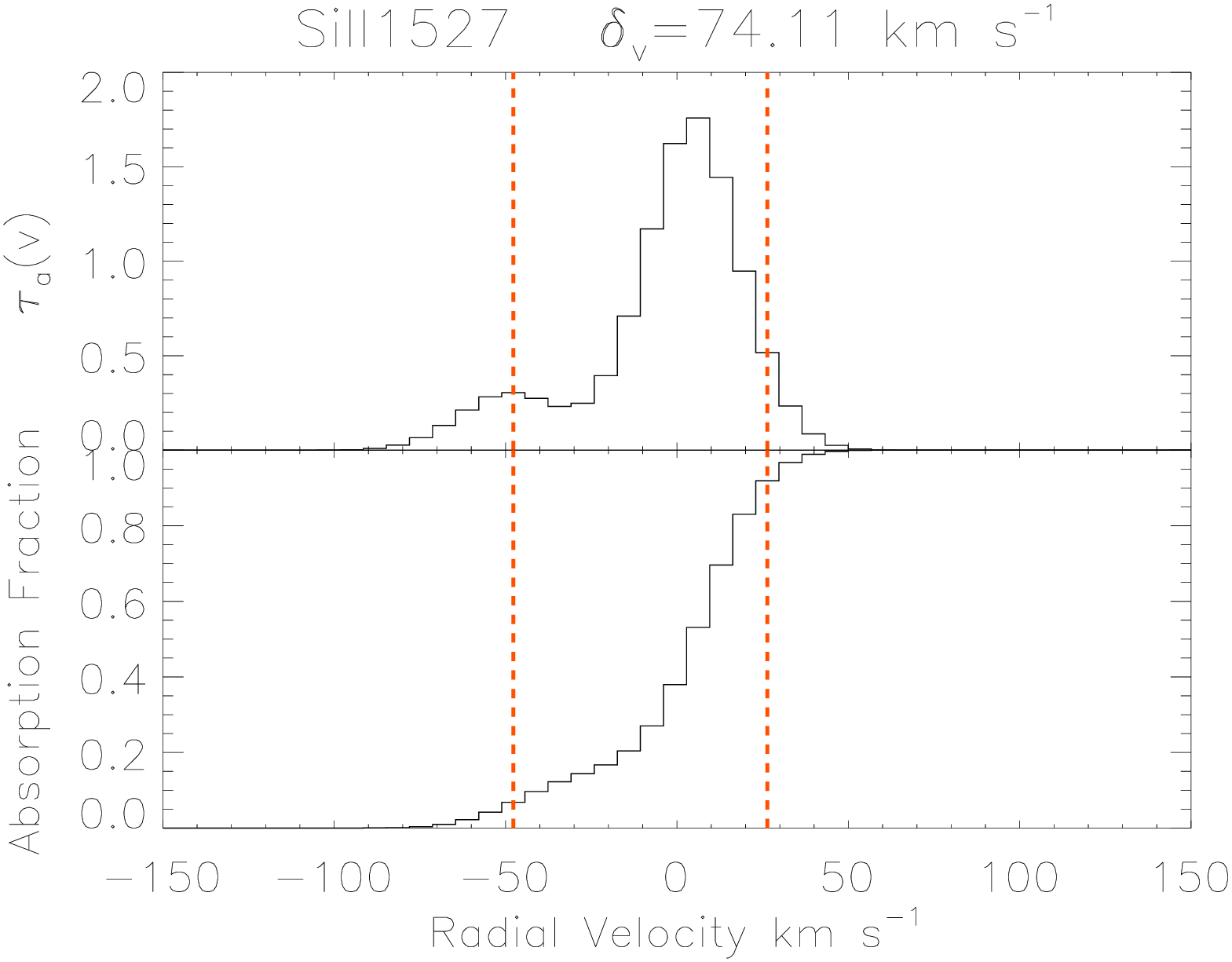}
\end{tabular}
\caption{Velocity dispersion for Si II $\lambda$1527 line in $z=4.829$ absorber toward Q0824+1302 based on our Voigt profile fits. The upper panel shows the optical depth profile and the lower panel shows the absorption fraction as a function of radial velocity. The 5 percent and 95 percent levels of the absorption are marked with two vertical dotted lines in red. }
\label{Fig:4829_kinematics}
\end{figure*}

Fig. \ref{Fig:4829_kinematics} shows the optical depth and the absorption fraction as a function of radial velocity for Si II $\lambda$1527 using the fitted profiles. The gas in this absorber is spread over a velocity width of 74 km s$^{-1}$. Using the metallicity vs. velocity dispersion relation [X/H] = $(1.52\pm0.08)$ log$\Delta V_{90} - (4.20\pm0.16)$ observed for lower redshift DLAs \citep[]{Quiret et al. 2016}, an absorber with $\Delta V_{90}$ = 74 km s$^{-1}$ would typically have a metallicity of -1.36 dex. This shows that the $z=4.829$ absorber has a metallicity far below the value predicted from the metallicity-velocity relation for DLAs. \citet{Morrison et al. 2016} found a similar velocity dispersion of  84 km s$^{-1}$ for a sub-DLA at $z=4.977$ that has a metallicity of -2.05 dex. 
\section{Discussion}
\label{sec:dust}
We now discuss the chemical and kinematic properties of the absorbers in our sample, and compare those with the properties of other absorbers from the literature at $z > 4.5$ \citep[e.g.][]{Rafelski et al. 2012, Rafelski et al. 2014, Morrison et al. 2016}. We note that all the absorbers are at velocity separations of $> 3000$ km s$^{-1}$ from the background quasars, and therefore, not likely to be associated with the quasars\footnote{We include the $z = 5.1791$ absorber toward Q1626+2751 in our sample, 
because it is not within the range of 3000 km s$^{-1}$ from the quasar emission redshift of 5.288 (the latter based on the SDSS Data Release 14).}. We also compare these $z > 4.5$ absorbers with those at $z < 4.5$ \citep[e.g.][and references therein]{Rafelski et al. 2012, Kulkarni et al. 2015, Som et al. 2015, Morrison et al. 2016, Quiret et al. 2016}. 
\subsection{Dust Depletion}
\begin{figure*}
\begin{tabular}{ll}
\includegraphics[scale=0.35]{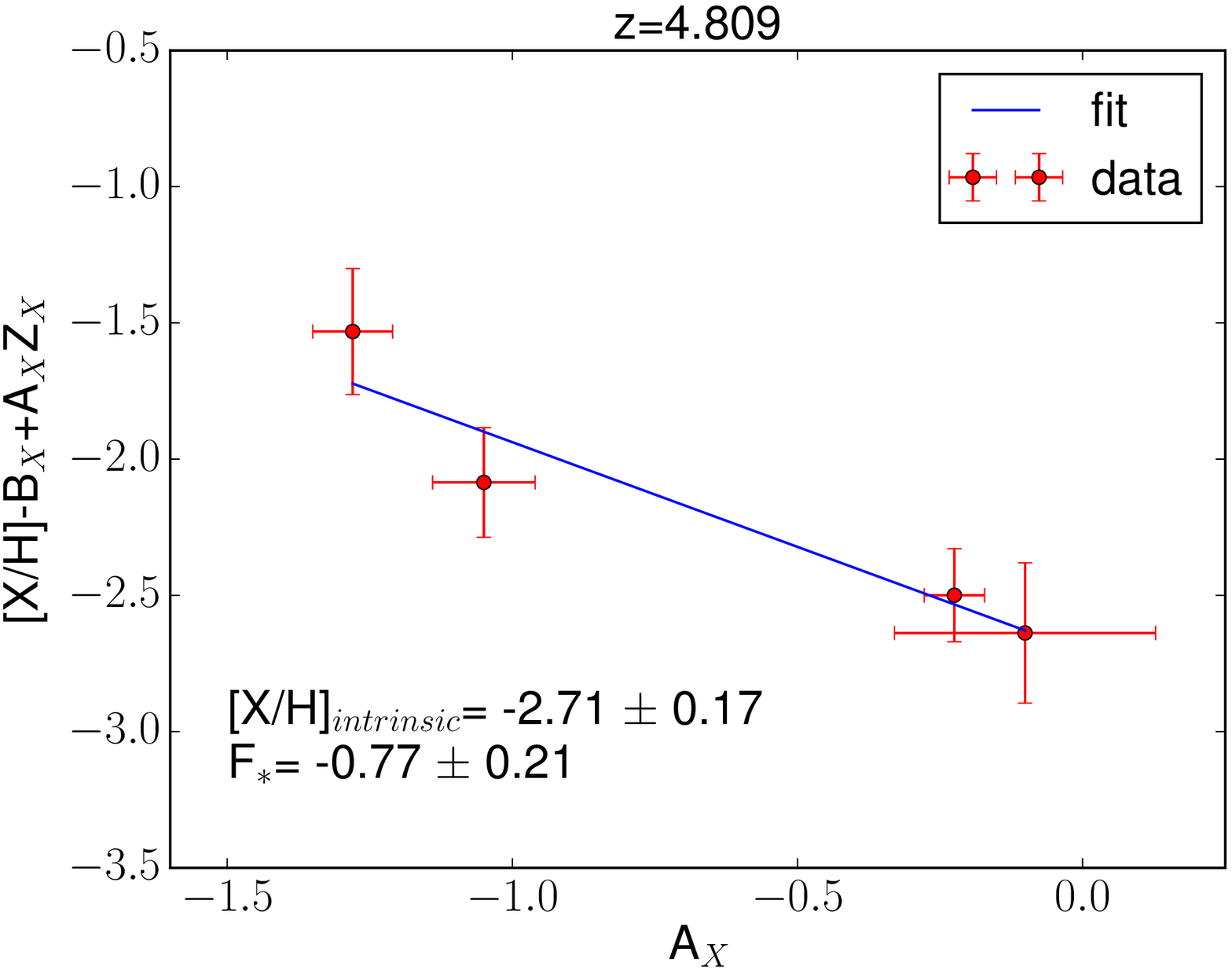}
\includegraphics[scale=0.35]{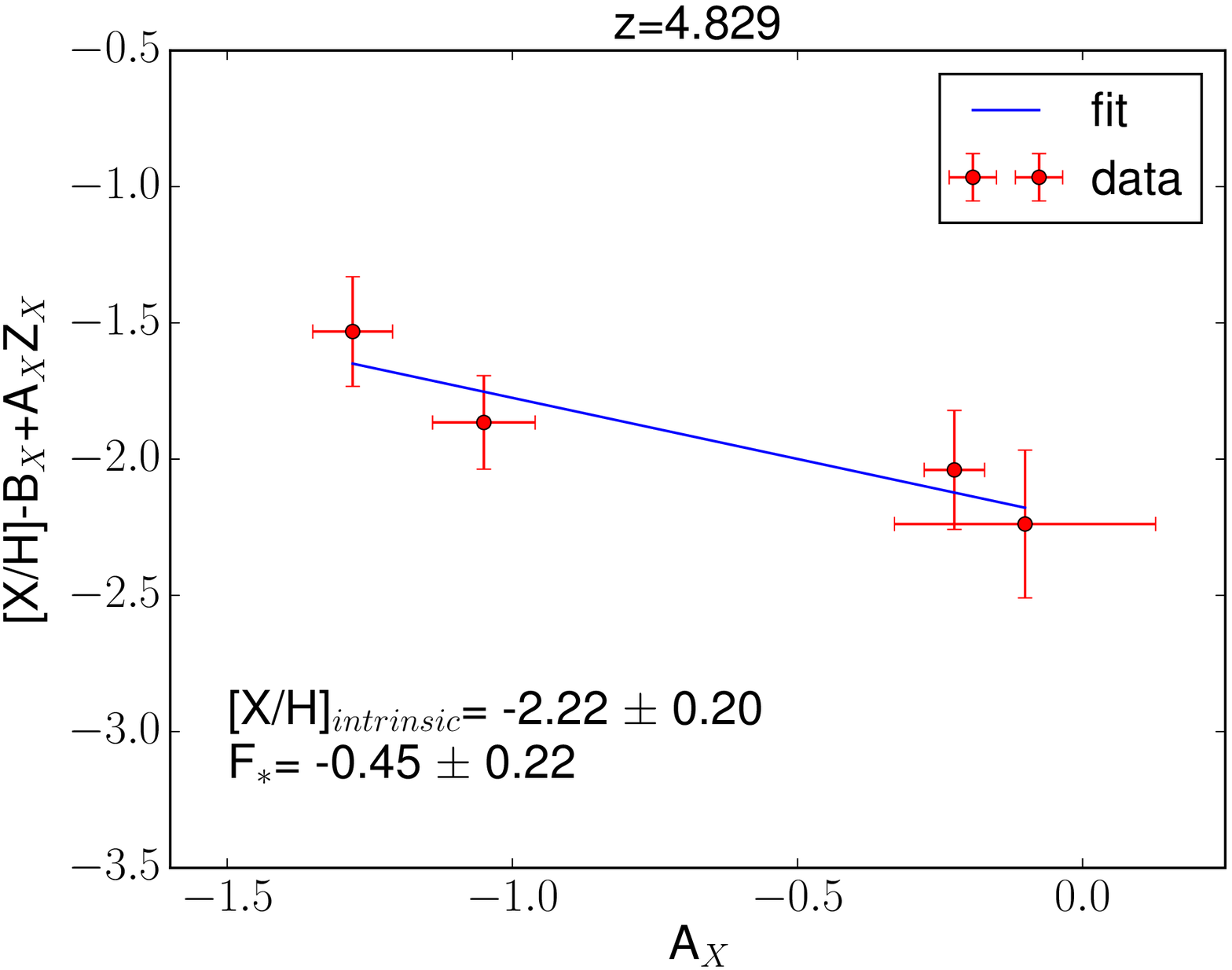}\\[2\tabcolsep]
\includegraphics[scale=0.35]{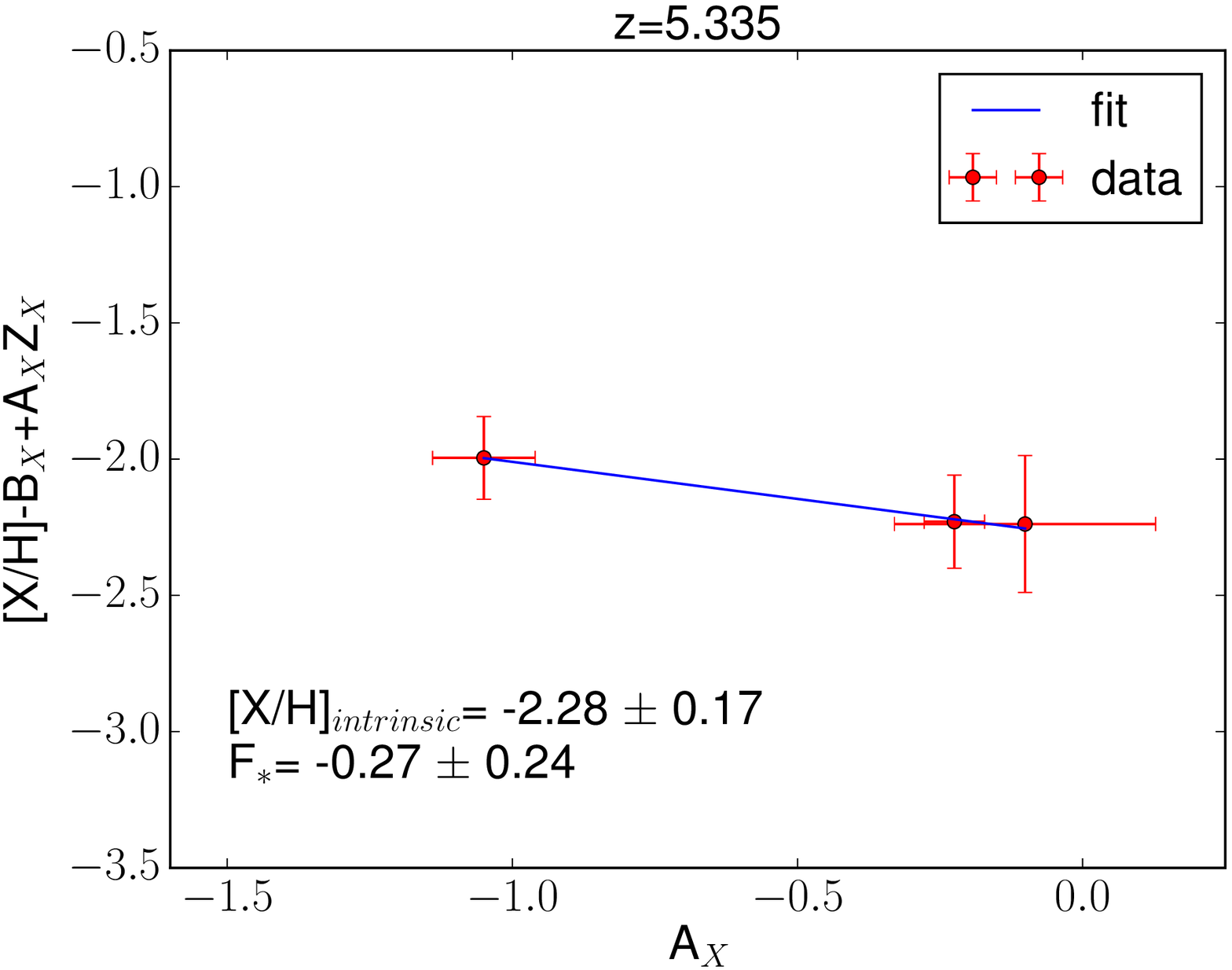}
\end{tabular}
\caption{Results of the Jenkin's approach to determine the intrinsic metallicity and $F_{*}$ value for the $z=4.809$, $z=4.829$ and $z=5.335$ systems. In each panel, the data points are shown in red and the regression fit to the data is shown by a blue line.}
\label{fig:jenkins}
\end{figure*}

Fig.~\ref{fig:jenkins} shows the results using the prescription of \citet{Jenkins 2009} to determine
the intrinsic metallicity and $F_{*}$ values for all the absorbers analyzed by us.
The $F_{*}$ values that exhibit the extent of dust depletion are
in the range of $-0.27\pm0.24$ to $-0.77\pm0.21$. The $F_{*}$ value for the $z$=4.977 absorber from \citet{Morrison et al. 2016} is $-0.14 \pm 0.17$. Overall, the four $z \sim 5$ absorbers with $F_{*}$ measurements [three from this work and one from \citet{Morrison et al. 2016}], have fairly negative $F_{*}$ values, as also found for lower redshift DLAs and sub-DLAs \citep[]{Quiret et al. 2016}. These $F_{*}$ values are much closer to the Milky Way's halo gas ($F_{*} = -0.28$) than to the Milky Way's cool disk gas, warm disk gas, or disk+halo gas ($F_{*} = 0.90, 0.12, $ and -0.08, respectively). \\

The relative abundances reported here, together with the measurements in the $z=4.977$ absorber from \citet{Morrison et al. 2016}, show several interesting indications of dust depletion and unusual nucleosynthesis. Three out of the four absorbers show significantly subsolar [C/O] ratios. The [C/O] ratios found in these $z\sim5$ absorbers are similar to the typical value of $-0.28\pm0.12$ for the very metal-poor (VMP) DLAs \citep[]{Cooke et al. 2011, Cooke et al. 2017}. Two out of four of our $z\sim5$ absorbers show strong depletion of Fe relative to O. One out of the four shows strong Si depletion relative to O. Two other absorbers show mild Si depletion, with [Si/O] comparable to the typical value of $-0.08\pm0.10$ for VMP DLAs \citep[]{Cooke et al. 2011}. The fourth absorber shows an excess of Si relative to O, which could be a leftover signature of nucleosynthesis by population III stars. Indeed, a substantial Si/O ratio can be expected in a  population III initial mass function (IMF) extending over 100-260 $M_{\odot}$ \citep[e.g.][]{Kulkarni et al. 2013}.\\

\begin{figure*}
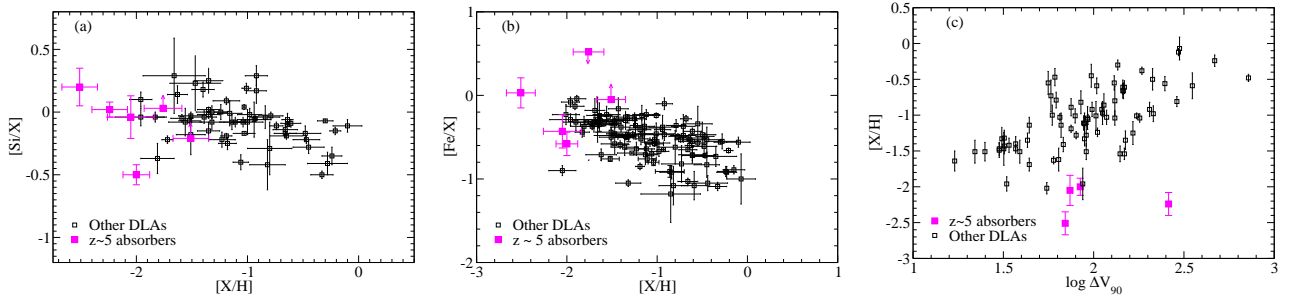

\begin{tabular}{ll}
\includegraphics[scale=0.22]{six_ref}
\hskip 0.1in
\includegraphics[scale=0.22]{fex_ref}
\hskip 0.1in
\includegraphics[scale=0.22]{deltav_ref}
\end{tabular}
\caption{(a) Plot of [Si/X] vs. metallicity [X/H] for $z \sim 5$ absorbers (shown with magenta squares) and lower-redshift DLAs from the literature (shown as black squares). (b) Plot of [Fe/X] vs. metallicity [X/H] for $z \sim 5$ absorbers (shown with magenta squares) and lower-redshift DLAs from the literature (shown as black squares). (c) Plot of metallicity [X/H] vs. velocity dispersion for $z \sim 5$ absorbers (shown in magenta) and lower-redshift DLAs (shown in black). In all panels, we use X=O for the $z \sim 5$ absorbers, and X=O, S, or Zn for the lower-redshift DLAs.}
\label{fig:six}
\end{figure*}

Fig.~\ref{fig:six}(a) shows a plot of depletion of Si with respect to undepleted elements O or Zn or S vs. metallicity. A correlation between depletion and metallicity has been noted before for lower-redshift DLAs \citep[e.g.][]{Meiring et al. 2006, Meiring et al. 2009, Noterdaeme et al. 2008, Kulkarni et al. 2015}. One of the $z \sim 5$ absorbers falls below this trend, while the others may be consistent with that trend. Fig.~\ref{fig:six}(b) shows a plot of depletion of Fe with respect to undepleted elements O or Zn or S vs. metallicity. Three of the $z \sim 5$ absorbers may be consistent with the correlation between depletion and metallicity for lower-redshift DLAs. Measurements of undepleted as well as depleted elements in a larger sample of high-z absorbers are needed to further clarify the nature of this trend at $z\ga5$.\\

\subsection{Metallicity vs. Velocity Dispersion Relation}
Fig.~\ref{fig:six}(c) shows a plot of metallicity vs. velocity dispersion for the $z \sim 5$ absorbers from this work and the literature, as well as lower redshift DLAs from the literature \citep[]{Kulkarni et al. 2015}. A strong correlation between metallicity and velocity dispersion has been reported for lower redshift DLAs and sub-DLAs \citep[e.g.][]{ Peroux et al. 2003, Ledoux et al. 2006, Moller et al. 2013, Som et al. 2015}. If the velocity dispersion is  taken as a measure of the mass of the absorber, such a correlation may imply a mass-metallicity relation. A correlation between the gas-phase metallicity and the stellar mass is indeed well-established in nearby and distant star-forming galaxies (SFGs) \citep[e.g.][]{Tremonti et al. 2004, Savaglio et al. 2005, Erb et al. 2006}. Furthermore, the mass-metallicity relation (MZR) shows evolution with redshift. \citet{Troncoso et al. 2014} report evolution in the MZR from $z = 3.5$ to $z= 0$, at a rate consistent with the overall rate of DLA metallicity evolution ($\sim$0.2 dex) at those redshifts. \citet{Shapley et al. 2017} report the first metallicity measurement at $z > 4$ in an SFG  based on nebular emission lines, and find a metalicity of about -0.77 dex, consistent with a drop of about 0.2 dex per unit redshift seen in lower redshift SFGs and DLAs. \\

Two of our $z\sim5$ systems fall distinctly below the metallicity-velocity trend for lower redshift absorbers, which can be understood as an evolution of the metallicity-velocity relation with redshift \citep[e.g.][]{Neeleman et al. 2013}. However, it is interesting to note that our $z \sim 5$ absorbers do not show much of a trend between metallicity and velocity dispersion. A much flatter metallicity vs. velocity dispersion relation at $z \sim 5$ compared to that at lower redshifts, if confirmed, would have interesting implications. For example, it may suggest that the $z \sim 5$ absorbers are tracing the more massive galaxies that did not get as enriched chemically; this may suggest that they have a higher M/L ratio and thus a lower $M_{*}$ and a lower metallicity. We note, however, that the velocity dispersion could partly arise in gas flows within the circumgalactic medium, and thus may not necessarily be a measure of the mass of the galaxy; in this scenario, the $z\sim5$ absorbers may trace galaxies with stronger inflows of less enriched material. Observations of more high-redshift absorbers are needed to establish whether these differences are unique to these particular absorbers or are a common phenomenon at these early epochs.

\subsection{Metallicity Evolution}
\begin{figure*}
\begin{tabular}{l}
\includegraphics[scale=0.45, clip]{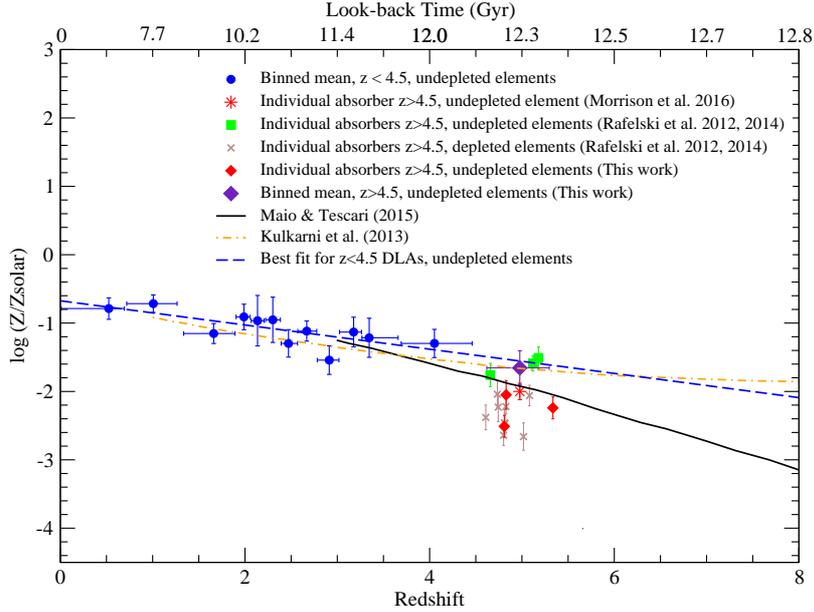}
\end{tabular}
\caption{Metallicity-redshift relation for DLAs. Blue circles show binned data for DLAs at $z < 4.5$ (with each bin containing 16 or 17 DLAs) based on undepleted elements \citep[]{Morrison et al. 2016}, and the dashed curve in blue shows the corresponding line of best fit. The green squares and brown crosses show previous measurements for individual DLAs at $z > 4.5$ based on elements that are undepleted in the Milky Way ISM, and those that are depleted, respectively \citep[]{Rafelski et al. 2012, Rafelski et al. 2014}. Red diamonds show results from this work based on the undepleted element O. Red star shows a sub-DLA from \citet{Morrison et al. 2016}. Indigo diamond shows the N$_{HI}$-weighted mean metallicity for undepleted elements at $z>4.5$. Solid curve shows the prediction for the mean gas metallicity from \citet*{Maio 2015}. The dotted-dashed curve in orange shows the prediction of the semi-analytic model of \citet{Kulkarni et al. 2013} including population II and population III stars.}
\label{fig:metallicity_evolution}
\end{figure*}

The metallicities of our $z \sim 5$ systems based on the undepleted element O are plotted in~Fig. \ref{fig:metallicity_evolution}. Also shown for comparison are the binned mean metallicities for DLAs at $z < 4.5$, and other measurements for individual absorbers at $z > 4.5$ from the literature. The solid curve shows the mean gas metallicity from the computations of \citet*{Maio 2015}, which do not include population III stars. The dotted-dashed curve shows the semi-analytic model of \citet{Kulkarni et al. 2013}, which includes population III  stars of 100-260 M$_{\odot}$, but is influenced mainly by population II stars.\\

Compared to predictions based on the metallicity-redshift relation for DLAs at $z < 4.5$ \citep[]{Morrison et al. 2016}, the metallicities of our systems at $z$ = 4.809, 5.335 and 4.829 are lower at the 4.1 $\sigma$, 2.5 $\sigma$, and 1.9 $\sigma$ levels, respectively. Combining these measurements with the others at $z > 4.5$ plotted in~Fig. \ref{fig:metallicity_evolution}, it is clear that 5 out of the 7 absorbers at $z > 4.5$ have metallicities [X/H] $\gtrsim$ -2.0.\\

\begin{figure*}
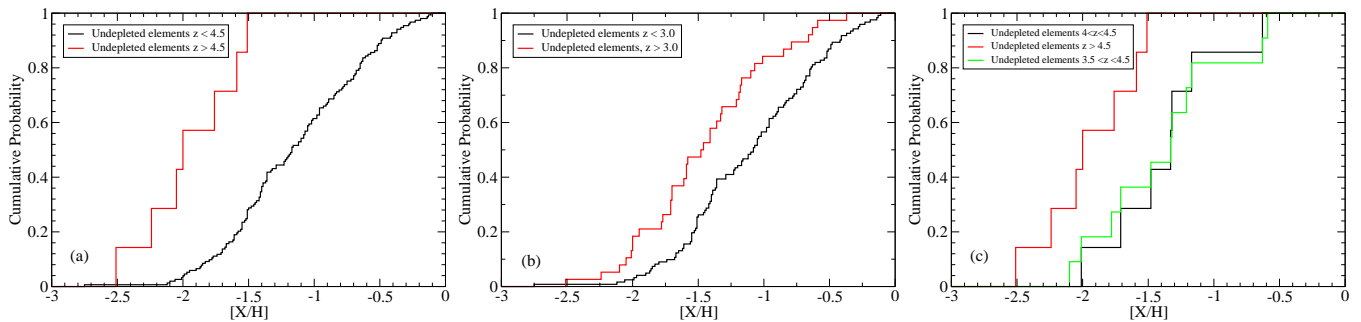

\begin{tabular}{ll}
\includegraphics[scale=0.24, clip]{undepleted4}
\includegraphics[scale=0.24, clip]{undepleted3}
\includegraphics[scale=0.24, clip]{ks_final}
\end{tabular}
\caption{a) A 2-sample K-S test comparing the cumulative probability functions of the metallicities based on undepleted elements for $z<4.5$ and $z>4.5$. b) Same as (a) but for $z<3$ and $z>3$. c) Same as (a) and (b), but for $z > 4.5$, $4 < z < 4.5$, and $3.5 < z < 4.5$.  The samples with $z < 4.5$ and $z > 4.5$ differ significantly, as do those with $z >3$ and $z <3$. However, the $z > 4.5$, $4 < z < 4.5$, and $3.5 < z < 4.5$ samples differ at 
$< 2 \sigma$ level.}
\label{fig:ks}
\end{figure*}

As a first step to examining whether the high and low redshift samples differ, we performed a 2-sample Kolmogorov-Smirnov (K-S) test, comparing the cumulative probability functions of the metallicities based on undepleted elements for $z < 4.5$ and $z > 4.5$, and found these two samples to be distinct (K-S test statistic D$_{\rm max} = 0.745$) at $> 3 \sigma$ significance level~(Fig. \ref{fig:ks} a). To explore this further, we next performed the 2-sample KS test on the absorbers with $z < 3$ and $z > 3$~(Fig. \ref{fig:ks}~b), and also for absorbers with $z < 3.5$ and $z > 3.5$. In each of these cases also, there is a $\sim 3 \sigma$ evidence for a difference in these populations (D$_{\rm max} = 0.312$ and D$_{\rm max} = 0.410$, respectively). This difference seems to arise from the fact that, in general, there is a gradual decrease in metallicity from low to high redshift (which is well-known and shown in~Fig. \ref{fig:metallicity_evolution}) and the $z<4.5$, $z<3.5$, or $z<3$ samples include absorbers spanning a very wide redshift range stretching down all the way to $z = 0$. Indeed, the range in look-back times at $0 < z < 4.5$ is much larger than that at $z > 4.5$. Thus, the results of the K-S test at redshifts above and below 4.5, by itself, is not necessarily an indication of a {\it sudden} drop in metallicity but rather a continuous decrease. To make the results of the K-S test less sensitive to the lowest redshifts (where the mean metallicity is the highest), we repeated the K-S test for the samples with $4 < z < 4.5$ and $z > 4.5$ (i.e., $4.5 < z < 5.4$). Both these samples have the same number of systems, and are free of the lower redshift metal-rich systems. The K-S test statistic for these two samples (D$_{\rm max} = 0.571$) is not significant at even the 2 $\sigma$ level, suggesting that there is no significant difference between the metallicities of undepleted elements in the absorbers at $4 < z < 4.5$ and those at $4.5 < z < 5.4$~(Fig. \ref{fig:ks} c). Likewise, there is no significant difference between the metallicities in the samples at $3.5<z<4.5$ and $4.5<z<5.4$~(Fig. \ref{fig:ks} c, D$_{\rm max} = 0.545$). 

To examine more quantitatively whether there is a drop in metallicity, we calculate the $N_{\rm H\, I}$-weighted mean metallicity 
for the 7 systems at $z > 4.5$ shown in~Fig. \ref{fig:metallicity_evolution} that have measurements of undepleted elements. Use of the $N_{\rm H\, I}$-weighted mean metallicity, which traces the global (cosmological) mean metallicity, i.e. the ratio of the mean co-moving densities of metals and hydrogen, $\Omega_{\rm metals}/\Omega_{\rm gas}$ \citep*[e.g.][]{Lanzetta 1995, Kulkarni 2002}, is standard practice in studies of DLA metallicity evolution \citep[e.g.][]{Prochaska et al. 2003a, Kulkarni et al. 2005, Kulkarni et al. 2010, Rafelski et al. 2012, Rafelski et al. 2014, Som et al. 2015, Quiret et al. 2016}. The $N_{\rm H\, I}$-weighted mean metallicity for the $z > 4.5$ systems with measurements of undepleted elements, shown in~Fig. \ref{fig:metallicity_evolution} as an indigo diamond, is $-1.66 \pm 0.25$ (where the uncertainty includes the sampling uncertainties as well as the measurement uncertainties in the metal and H I column densities). At the median redshift of this high-$z$ bin, i.e. $z = 4.98$, the predicted mean metallicity based on the metallicity-redshift relation used in \citet{Morrison et al. 2016} for DLAs at $z < 4.5$ is $-1.55 \pm 0.18$. Even ignoring the uncertainty in the predicted value, the 
difference between the observed and predicted $N_{\rm H\, I}$-weighted mean metallicity values is $< 0.5 \sigma$ significant. 
We thus conclude that there is no evidence of a sudden drop in metallicity at $z > 4.5$. The primary reason our finding differs from that of \citet{Rafelski et al. 2012, Rafelski et al. 2014} appears to lie in the fact that we are focusing on elements that do not deplete on dust grains. Observations of undepleted elements in more absorbers are essential to further test whether or not there is a drop in metallicity.

We emphasize in this context that the $N_{\rm H\, I}$-weighted mean metallicity derived here is dominated by absorbers with 
the higher H I column densities. Thus the inclusion of some absorbers at the boundary between DLAs and sub-DLAs in our sample has little impact on the $N_{\rm H\, I}$-weighted mean metallicity derived. We also note that our use of O I,  which is not sensitive to ionization corrections, 
means that the higher mean metallicity derived by us here is not an artifact of ionization.

Finally, we estimate the metallicity predicted at $z=5.335$, the highest redshift in our sample. Using the $N_{\rm H\, I}$-weighted mean metallicity in the highest redshift bin at $z <4.5$ and the value of -2.03 dex at $z=4.85$ from \citet{Rafelski et al. 2014}  to estimate an approximate "local slope", and extrapolating this relation further, one would expect the mean metallicity at $z=5.335$ to be -3.04 dex, if there was in fact a sudden drop in mean metallicity. Our observation of a metallicity of $-2.24 \pm 0.16$ dex in the $z=5.335$ absorber in our sample would make this absorber 5.0~$\sigma$ deviant from the expected mean at that redshift. While this is not entirely unlikely, the possibility of finding such an extreme value in the only $z > 5.3$ absorber explored seems low. 

A sudden drop in metallicity at $z > 4.7$ would also be surprising given that there is no evidence of a sudden change in the star formation history of the universe at redshifts leading up to $z \approx 4.7$. Given the absence of such a drop in our larger sample based on undepleted elements, we suggest that such a sudden drop may not be present. At the same time, we note that a sample of 7 systems is still too small to draw firm conclusions, and emphasize the need to obtain more measurements for undepleted elements at $z > 4.5$.

The metallicities of some of the $z \sim 5$ absorbers appear to be more consistent with the \citet*{Maio 2015} model, which does not include population III stars. These observed metallicities are substantially lower than those predicted by the \citet{Kulkarni et al. 2013} model. But as pointed by \citet{Kulkarni et al. 2014}, the rate of metal enrichment depends on the delay in the chemical enrichment of the gas by supernovae. The difference between some of the $z \sim 5$ DLAs and the \citet{Kulkarni et al. 2013} model could be a result of a longer delay in the supernova-driven enrichment than assumed in the model.\\

\section{Conclusions}
\label{sec:conclusion}
We have measured element abundances in three $z > 4.8$ absorption-selected gas-rich galaxies. These measurements have roughly doubled the existing sample of measurements for undepleted elements in DLAs/sub-DLAs at $z > 4.5$. Our main results are as follows: 

1. The $z \sim 5$ absorbers generally show low [C/O] ratios. The [C/O] ratios, and in some cases the [Si/O] ratios, are comparable to those in the VMP DLAs.

2. Based on the measurements of [Fe/O] and [Si/O], depletion of metals into dust grains appears to be prevalent in some $z \sim 5$ absorbers. Some absorbers lie below the depletion vs. metallicity trend  observed in the lower-redshift absorbers, but others may be consistent with that trend.

3. The current data show hints that the metallicity vs. velocity dispersion relation for $z \sim 5$ absorbers may be different from that for lower-redshift absorbers, although further measurements are needed to establish this 
more conclusively.

4. The $N_{\rm H\, I}$-weighted mean metallicity of $z\sim5$ DLAs appears to be consistent with the prediction based on $z<4.5$ DLAs.  Thus, our data do not suggest a sudden drop of metallicity at $z > 4.7$, but rather a continuous decrease. Measurements of undepleted elements in many more high-redshift absorbers are needed to definitively determine any changes in the metallicity evolution trend at $z > 4.7$. \\

Measurements of undepleted elements in a few dozen high-$z$ DLA/sub-DLAs would help to definitively determine the evolution of metallicity  and the velocity dispersion vs. metallicity relation in the first $\sim$1 Gyr of cosmic history. Measurements of both refractory (depleted) and volatile (undepleted) elements in the same absorbers are needed to determine the evolution of dust in this early epoch. More measurements of relative abundances, especially  [Si/O],  would also be useful to constrain  parameters of early  nucleosynthesis models, e.g. the population III IMF. Comparison of the trends resulting from these studies with predictions of cosmic chemical evolution models \citep*[e.g.][]{Pei et al. 1999, Somerville et al. 2001, Maio 2015} would offer invaluable new insights into the early history of star formation and chemical enrichment in galaxies.\\

To continue to understand the full  significance of high-$z$ DLA/sub-DLAs to SFGs, it will also be important to compare the chemical and kinematic properties of the absorbers at $z \gtrsim 5$ with SFGs at similar redshifts. While current SFG metallicity measurements extend only to $z \sim 4$, the James Webb Space Telescope is expected to allow metallicity measurements for SFGs out to $z \sim 10$ \citep[e.g.][]{Shapley et al. 2017}. It will be of great  interest  to compare the MZR and the redshift  evolution of the MZR for SFGs with the corresponding  quantities for absorption-selected galaxies such as those studied here.

\section*{Acknowledgements:}

We thank an anonymous referee for helpful comments that have improved this paper. SP and VPK acknowledge partial support from NSF grant AST/1108830, NASA grant NNX14AG74G and NASA/STScI support for HST programs GO-12536, 13801 (PI Kulkarni). This research has made use of the Keck Observatory Archive (KOA), which is operated by the W. M. Keck Observatory and the NASA Exoplanet Science Institute (NExScI), under contract with the National Aeronautics and Space Administration. 











\bsp	
\label{lastpage}
\end{document}